\documentclass[10pt,letterpaper]{article}

\usepackage[UKenglish]{babel}
\usepackage[margin=1in]{geometry}
\usepackage{microtype}
\usepackage{xcolor}
\usepackage{amsmath}
\usepackage{amsthm}
\usepackage[hidelinks]{hyperref}
\usepackage{ourmacros}
\usepackage{thm-restate}
\usepackage[capitalise,noabbrev]{cleveref}

\theoremstyle{plain}
\newtheorem{theorem}{Theorem}
\newtheorem{lemma}[theorem]{Lemma}
\newtheorem{corollary}[theorem]{Corollary}
\theoremstyle{definition}
\newtheorem{definition}[theorem]{Definition}
\newtheorem{example}[theorem]{Example}
\theoremstyle{remark}
\newtheorem{remark}[theorem]{Remark}

\graphicspath{{figures/}}
\title{An Omnidirectional WECT: Closed-Form Computation of the Directionally Integrated Weighted Euler Characteristic Transform}

\author{%
    Jessi Cisewski-Kehe\thanks{Department of Statistics, University of Wisconsin, Madison, Wisconsin, U.S.A;
        \url{https://www.jjckehe.com/}; \href{mailto:jjkehe@wisc.edu}{\nolinkurl{jjkehe@wisc.edu}};
        \url{https://orcid.org/0000-0002-9656-2272}; NSF grants 2337243 and 2346683.}
    \quad Brittany Terese Fasy\thanks{Montana State University, Bozeman, Montana, U.S.A.
        Brittany Terese Fasy: \url{https://www.fasy.us/};
        \href{mailto:brittany.fasy@montana.edu}{\nolinkurl{brittany.fasy@montana.edu}};
        \url{https://orcid.org/0000-0003-1908-0154}; NSF grants 2046730 and 2152737.
        Alexander McCleary: \url{https://www.alexmccleary.org/};
        \href{mailto:alex.mccleary@gmail.com}{\nolinkurl{alex.mccleary@gmail.com}};
        \url{https://orcid.org/0000-0002-2423-7518}; NSF grants 2046730 and 2152737.}
    \quad Alexander McCleary\footnotemark[2]
    \quad Eli Quist\thanks{TeraDact, Missoula, Montana, U.S.A;
        \url{https://eliquist.xyz/}; \href{mailto:quistew@gmail.com}{\nolinkurl{quistew@gmail.com}};
        \url{https://orcid.org/0009-0004-5878-9718}; NSF grants 2046730 and 2152737.}\\
    Jacob Sriraman\footnotemark[2]}
\date{}
\hypersetup{
    pdftitle={An Omnidirectional WECT: Closed-Form Computation of the Directionally Integrated Weighted Euler Characteristic Transform},
    pdfauthor={Jessi Cisewski-Kehe; Brittany Terese Fasy; Alexander McCleary; Eli Quist; Jacob Sriraman}
}

\begin{document}

\maketitle

\begin{abstract}
    The weighted Euler characteristic transform (WECT) and related directional
transforms have been used for shape analysis across a variety of domains.
In practice, however, they are approximated over a finite set of filtration directions,
and thus the vectorized descriptor depends on this choice of directions.
In this manuscript, we formally introduce the directionally integrated weighted
Euler characteristic transform (DIWECT), which averages the WECT over all directions~$s \in \sph^n$ for a weighted simplicial complex immersed in~$\R^{n+1}$.
We provide a framework for computing the DIWECT exactly, without sampling
directions.
We provide implementations for complexes immersed in~$\R^2$ and~$\R^3$,
where the computation reduces to computing arc lengths and spherical areas,
respectively.
Experiments on synthetic and real-world data show that the DIWECT,
while resulting in more compact vectorizations, also successfully
encodes shape information and removes the need for~alignment.

\end{abstract}

\section{Introduction}

The weighted Euler characteristic transform (WECT) \cite{Turner2014PHT} is an
instrumental tool in the field of topological data analysis.
The WECT and its variants have proven to be a powerful shape invariant
in many applied settings, including image analysis
\cite{Jiang2020TheWE, cisewski2023weighted},
molecule shape analysis \cite{ToscanoDuran2026},
the time evolution of organoids \cite{marsh2024detecting},
and simulating atomized sprays \cite{bubbles}.
Its utility is largely because it is a data summary rich in topological information
that is easy to vectorize, making it straightforward to
incorporate into statistical and machine learning~pipelines.

The WECT of a shape immersed in~$\R^{n+1}$ is a family of step functions, one
for each direction
in~$\sph^n$, where each step function
tracks the shape's
weighted Euler characteristic across a height filtration.
Hence, the WECT is parameterized by two components:
the chosen direction, and the corresponding height parameter.
Recent work has focused on efficient and optimized computation of the WECT
and its variants
\cite{SoCGpyECT, fasttopology, eucalc, wang2023gpu_conference, wang2023gpu, ayub2026ect}.

While computing the WECT has become more efficient, analysis challenges
exist when comparing WECTs of different datasets due to limitations
in aligning the directions, as is typical in classification or prediction tasks.
Furthermore, computing a high-resolution vectorization of the WECT
(with respect to the number of sampled directions and heights)
leads to a massively high-dimensional representation.
In this work, we offer a computationally efficient alternative method that
removes the dependency on the direction parameter by integrating over all
possible directions of height filtrations, called the
{\it directionally integrated WECT} (DIWECT).
The initial formulation of the DIWECT was applied to address challenges with
shape registration in a matching task used to simulate
fluid atomization \cite{bubbles}.
Here, we define and characterize the theoretical foundations of the DIWECT, and offer
a fast computational framework for computing it in~$\mathbb{R}^2$ and~$\mathbb{R}^3$.

Previous work on the DIWECT approximated the integral over the direction parameter \cite{bubbles}.
In particular, they estimate the DIWECT by first computing
the WECT using a fixed, uniformly sampled set of directions,
and then computing an average for each height parameter
over this set of directions.
Here, we avoid this approximation and propose a fast and direct implementation of the DIWECT.
Another related framework, \cite{marsh2024detecting}, incorporates directional averaging
to achieve rotational invariance.
However, the pipeline does not utilize weighted complexes,
does not specify how the integral over the sphere is computed in practice,
and is focused on time-series representations of data.

This manuscript reduces the problem of integrating over all directions
to summing exactly computed areas of regions on the sphere.
Sampling the DIWECT across the filtration height parameter creates a rotationally
invariant, compact, and quickly computable topological
descriptor that may be used as input to downstream analysis and learning tasks.
We demonstrate the utility of this descriptor on a variety of synthetic and
real-world datasets.

\section{Background}

\subsection{The Weighted Euler Characteristic Transform}

We adopt the following notation and definitions
used in \cite{SoCGpyECT}.
Given a simplicial complex~$K$, we denote the vertex set of~$K$ as~$K_0$, and the set of subcomplexes of~$K$ as~$\sub(K)$.

\begin{definition}[Weighted Simplicial Complex]
    A \emph{weighted simplicial complex} is a simplicial complex~$K$ and
    a weight function~$\omega \colon K \to \R$, written as a pair~$\K = (K,\omega)$. 
\end{definition}

Throughout the paper,~$\K$ denotes a weighted simplicial complex and~$K$ its
underlying simplicial complex.
In particular, we write~$\sigma \in K$ for a simplex of~$\K$,
and subscript transforms of~$\K$ (such as~$\wect_\K$ and~$\diwect_\K$) by~$\K$.
For example, consider starting with a grayscale image, and let the pixel
intensities define vertex weights that can then propagate to higher-dimensional
simplexes in a Freudenthal triangulation~\cite{Jiang2020TheWE, cisewski2023weighted}.

Topological invariants summarize the weighted simplicial complex; for example,
consider the following one-number summary:

\begin{definition}[Weighted Euler Characteristic]
    The \emph{weighted Euler characteristic~(WEC)} is
    \[ \chi(\K) := \sum_{\sigma \in K} (-1)^{\dim{\sigma}} \omega(\sigma), \]
    where~$\K = (K, \omega)$ is a weighted simplicial complex.
\end{definition}

As one number is often insufficient to describe a space, we turn to filtered
complexes. In particular, we filter with respect to a direction~$s \in \S^n$:

\begin{definition}[Directional Filtration]
    Let~$K$ be a simplicial complex immersed in~$\R^{n+1}$,
    and~$\vect{s} \in \sph^{n}$ a direction vector.
    The \emph{height function} of a simplex~$\sigma \in K$
    is the function~$\heightfcn{\sigma} \colon \sph^n \to \R$ defined as~$\height{\sigma}{\vect{s}} := \max_{v \in \sigma} v \cdot \vect{s}$.
    The \emph{directional filtration}
    in direction~$\vect{s}$ is the function~$H_{\vect{s}} \colon \R \to \sub(K)$, defined as~$H_{\vect{s}}(t) := \{ \sigma \in K \mid \height{\sigma}{\vect{s}} \leq t \}. $
\end{definition}

For a simplex~$\sigma \in K$,
let~$\ifn_{\sigma} : \sph^{n} \times \R \to \{0, 1\}$
be the indicator function denoting whether~$\sigma$ is included in a
height filtration for a given direction at a given height:
\begin{equation}\label{eq:indicator}
    \ifn_{\sigma}(\vect{s}, t) = \begin{cases}
        1 \quad \text{if } \sigma \in H_{\vect{s}}(t), \\
        0 \quad \text{otherwise}.
    \end{cases}
\end{equation}

Tracking the WEC throughout height filtrations in all possible directions yields
the~WECT.

\begin{definition}[Weighted Euler Characteristic Transform]\label{def:wect}
    Given weighted simplicial complex~$\K = (K, \omega)$ immersed in~$\R^{n+1}$,  the function~$\wect_{\K} \colon \sph^{n} \times \R \to \R$ is the weighted Euler characteristic transform (WECT) of~$\K$ defined as
    \begin{equation}
        \wect_{\K}(\vect{s}, t) := \chi \big( H_{\vect{s}} (t), \omega|_{H_{\vect{s}}(t)} \big)
        = \sum_{\sigma \in K} (-1)^{\dim{\sigma}} \omega(\sigma) \ifn_\sigma(\vect s, t).
    \end{equation}
\end{definition}

The WECT is a \emph{faithful descriptor}:
two simplicial complexes immersed in~$\R^{n+1}$ have the same WECT if and only
if they are the same complex and~immersion~\cite{fasy2025faithful}. We note that, in some related work,
this is called a \emph{sufficient statistic}, a \emph{complete invariant}, or
\emph{injective}.

\begin{remark}
    A widely-used special case of the WECT is the Euler
    characteristic transform (ECT)~\cite{Turner2014PHT} where~$\omega(\sigma) = 1$ for each~$\sigma \in K$,
    with its corresponding Euler characteristic function (ECF) defined analogously.
\end{remark}

\subsection{Related Work}

The persistent homology transform (PHT) was introduced in \cite{Turner2014PHT}
alongside the ECT. The PHT is similar to the ECT, but persistence diagrams are
computed in each direction instead of ECFs.  The ECT and WECT have been used in various data analysis settings:
barley seed classification~\cite{amezquita2022measuring}, analysis of cancer imaging data~\cite{mohammed2024statistical},
and
3D shape reconstruction~\cite{nadimpalli2023euler}. Alignment is a preprocessing step that is often tailored to a dataset, though some have proposed general methods
(e.g.,~\cite{kirveslahti2024digital}).  Another option is to integrate across all
directions: to avoid the alignment step and decrease the size of the descriptor, \cite{bubbles} and \cite{marsh2024detecting} average the
WECT~(or a variant of the WECT) over a finite number of directions.

In \cite{bubbles}, the vectorized WECT is averaged over~$150$ directions to produce a data descriptor for
droplets in three-dimensional atomized sprays.
This is then used as input to distance-based clustering and
shape-matching tasks on the droplet data.
Here, the rotational invariance is important, as the alignment of the atomized shapes is not fixed.
This work highlights the tradeoff between meeting computational resource limitations
while still encoding sufficient information of each shape and achieving
theoretical guarantees.

In \cite{marsh2024detecting}, the {\it Detecting Temporal shape changes with the
Euler Characteristic Transform} (DETECT) is introduced
to perform a temporal analysis of mouse organoids.
DETECT is based on the smooth ECT
(SECT)~\cite{crawford2020predicting}.
The SECT is determined by the same
family of height filtrations as the WECT, but centers each \emph{unweighted} ECF and
smooths it by integrating to the cumulative ECF to obtain a piecewise linear
function (as opposed to a step function).
DETECT then integrates the SECT across all directions.
The resulting descriptors are tracked through a temporal parameter.
However, \cite{marsh2024detecting} does not specify how this integral over the
sphere is computed or approximated.

\section{DIWECT: Definition and Computation} \label{sec:diwect}

In this section, we define the
directionally integrated weighted Euler characteristic transform (DIWECT)
and prove its closed-form formulas for
complexes immersed in~$\R^2$ and~$\R^3$.

\subsection{Definition}

The WECT has two parameters: a direction~$\vect s$ and a height~$t$.
The DIWECT
removes the dependence on direction by integrating over
all possible directions of the height filtration.
\begin{definition}[DIWECT]
    Let~$\K= (K, \omega)$ be a weighted simplicial complex immersed in~$\R^{n+1}$.
    The directionally integrated weighted Euler characteristic transform
    (DIWECT) of~$\K$ is the function~$\diwect_\K: \R \to \R$ defined by integrating the WECT over all directions:
   \begin{equation} \label{eq:diwect}
        \diwect_\K(t)
        := \int_{\vect{s} \in \sph^n} \wect_\K( \vect{s}, t) \dif \vect{s}.
    \end{equation}
\end{definition}
We note here that, unlike the WECT, the DIWECT is not a faithful descriptor:

\begin{example}[DIWECT is Not Faithful]
    Consider the following two complexes, each with two points:~$K=\{v_1=(1,0),v_2=(\sqrt{2},\sqrt{2})\}$ and~$L=\{w_1=(1,0),w_2=(0,2)\}$.
    We note that~$\diwect_{\K}(t) = \diwect_{v_1}(t)+\diwect_{v_2}(t)$ and~$\diwect_{\L}(t)= \diwect_{w_1}(t)+\diwect_{w_2}(t)$.  Moreover,
    because~$w_2$ is a rotation of~$v_2$ about the origin,~$\diwect_{v_2}(t)=\diwect_{w_2}(t)$.
    In addition, noting that~$v_1=w_1$, we get~$\diwect_{\K}(t)=\diwect_{\L}(t)$.
\end{example}

\subsection{Computation}

Expanding integral in \eqref{diwect} by using \defref{wect} (WECT),
we~obtain
\begin{align}
    \diwect_\K(t)
    &= \int_{\vect s \in \sph^n} \bigg( \sum_{\sigma \in K} (-1)^{\dim{\sigma}} \omega(\sigma) \ifn_\sigma(\vect s, t) \bigg) \dif \vect s \notag \\
    &= \sum_{\sigma \in K} \bigg( (-1)^{\dim(\sigma)} \omega(\sigma)
        \int_{\vect s \in \sph^n}\ifn_\sigma(\vect s, t) \dif \vect s \bigg).\label{eq:diwectint}
\end{align}
Therefore, to solve the integral in \eqref{diwect}, it suffices to solve the
integral in \eqref{diwectint}
for each~$\sigma \in K$.
The integral of an indicator function is, by definition, the Hausdorff measure
(denoted~$\mu$)
of the support of that indicator function; that is, we must compute surface
areas of subsets of~$\S^n$.
Unless otherwise specified,~$\mu$ will implicitly match the dimension of~$\sph^n$ in context (e.g., one-dimensional length for~$\sph^1$ and two-dimensional
surface area for~$\sph^2$).

\begin{definition}[Support of a Simplex]
    Let~$\sigma$ be a geometric simplex in~$\R^{n+1}$.
    For each~$t \in \R$, define the support of~$\sigma$ at height~$t$,
    denoted~$\suppfcn{\sigma}{t} \subseteq \sph^n$, as
    \begin{equation} \label{def:A-sigma}
        \suppfcn{\sigma}{t} := \supp\left(\ifn_\sigma(\vect s, t)\right) = \{ \vect s \in \sph^n \mid \height{\sigma}{\vect s} \leq
        t \}.
    \end{equation}
\end{definition}
That is,~$\suppfcn{\sigma}{t}$ is the set of directions for which~$\sigma$ has
height at most~$t$.
Fixing a height~$t \in \R$, the integral in \eqref{ind-fn-integral} is equivalent to the
measure of~$\suppfcn{\sigma}{t}$:
\begin{equation} \label{eq:ind-fn-integral}
    \int_{\vect s \in \sph^n} \ifn_\sigma(\vect s,t) \dif \vect s = \mu \big(
    \suppfcn{\sigma}{t} \big).
\end{equation}

\begin{figure}
    \centering
    \includegraphics[scale=0.5]{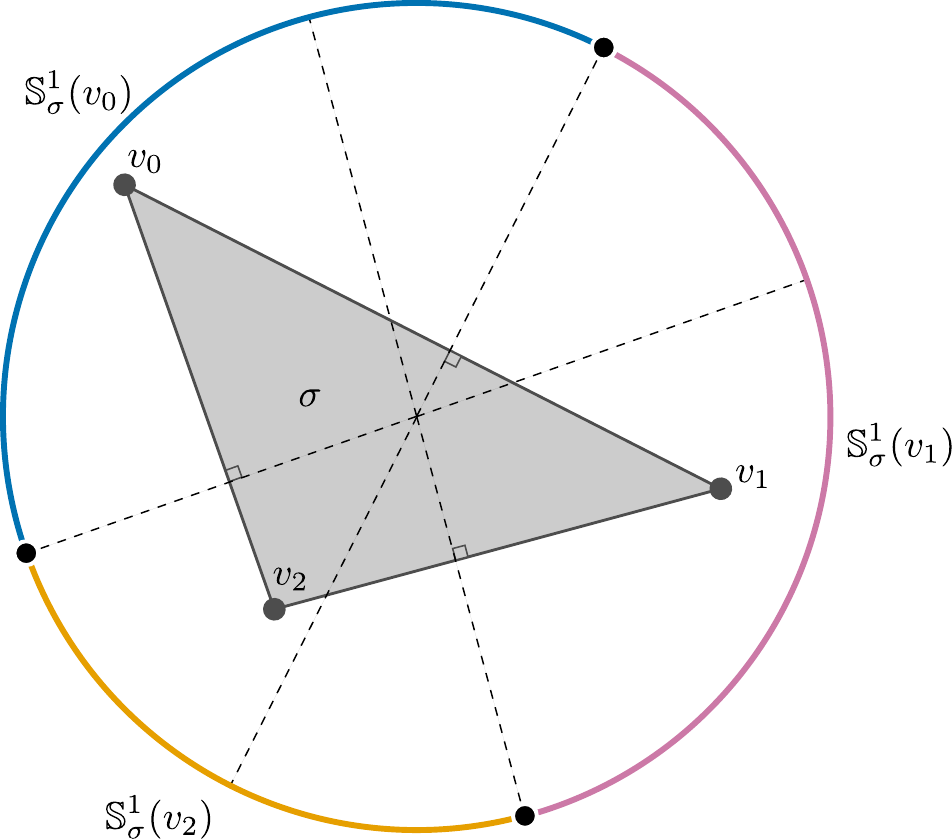}
    \caption{
        The subsets~$\vwit{1}{\sigma}{\vect{v_i}}$ for the two-simplex~$\sigma$ (orange, pink, and blue directions for~$i=0, 1,$ and~$2$, respectively);
        see Definition \ref{def:sph-sigma}. The black lines indicate the
    directions of~$\sph^1$ for multiple vertices tie. The three black points
    represent such directions and compose~$T_\sigma$.}
        \label{fig:S1-sigma}
\end{figure}

To compute~$\mu \big(\suppfcn{\sigma}{t} \big)$,
we decompose it according to which vertex
attains the maximum height. We call these sets
\emph{vertex witness regions}, which are defined as follows:
\begin{definition}[Vertex Witness Region] \label{def:sph-sigma}
    Let~$\sigma$ be a simplex in~$\R^{n+1}$.
    For each vertex~$\vect v \in \sigma$, the \emph{vertex witness region} of~$\vect v$ is the subset~$\vwit{n}{\sigma}{\vect{v}} \subseteq \sph^n$ for which~$\vect v$ is the (unique) highest vertex of~$\sigma$. That is,
    \begin{equation}
        \vwit{n}{\sigma}{\vect{v}}
            = \left\{ \vect s \in \sph^n \mid
                    \forall \vect w \neq \vect{v} \in \sigma,\,
                        \height{\vect w}{\vect s} < \height{\vect v}{\vect{s}}
              \right\}.
    \end{equation}
\end{definition}
\figref{S1-sigma} provides an example of the three vertex witness regions for a
two-simplex in~$\R^2$.
Geometrically,~$\vwit{n}{\sigma}{\vect{v}}$ is the intersection of open hemispheres
in~$\S^n$, and is hence~$n$-dimensional.
The closure of the vertex witness region is denoted~$\closedvwit{n}{\sigma}{\vect v}$.
\journal{And, the set~$\S^n \setminus
\{\vwit{n}{\sigma}{\vect{v}}\}_{v \in \sigma}$ is the set of directions for which two or more vertices
of~$\sigma$ share the same height as~$\sigma$; this set has Hausdorff dimension
strictly less than~$n$.
}

\begin{lemma}[Partition by Vertex Witness Regions] \label{lem:partition}
    Given a geometric~$k$-simplex~$\sigma$ in~$\R^{n+1}$, the
    vertex witness regions~$\{\vwit{n}{\sigma}{\vect v}\}_{\vect v \in \sigma}$
    are disjoint and~$\mu(\S^n)=\mu(\bigcup_{\vect v \in
    \sigma}\vwit{n}{\sigma}{\vect v})$.
\end{lemma}

\begin{proof}
    The fact that the vertex witness regions are disjoint follows from the
    definition.
    For the second part of the proof, let~$\tau \preceq \sigma$ such that~$\dim(\tau)>0$. We define~$T_{\tau}$ to be the set of directions for which
    the vertices of~$\tau$ are all at the same height as~$\sigma$, and all other
    vertices are below the vertices of~$\tau$.
    By construction, the sets~$\{\vwit{n}{\sigma}{\vect v} \}_{\vect v \in \sigma}$
    and~$\{T_{\tau}\}_{\tau \preceq \sigma, \dim(\tau)>0}$ partition~$\S^n$.
    Let~$\vect{v},\vect{w}$ be two distinct vertices in~$\tau$. Letting~$\vect{n}$
    be a vector normal to the line passing through~$\vect{v}$ and~$\vect{w}$ and~$H$ be the hyperplane defined by the normal~$\vect{n}$ and passing through the
    origin, we know that~$T_{\tau} \subset H \cap \S^n$.  Observing that the Hausdorff dimension of~$H
    \cap S^n$ is~$n-1$, we conclude the proof.
\end{proof}

Next, we build up the mechanisms required to measure the contribution to the
DIWECT of each vertex witness region for a simplex at a given filtration value~$t$. As such, we first define a function to normalize a vertex onto the surface of~$\sph^n$.

\begin{definition}[Normalized Vector]
    \label{def:normalized-vector}
    Define~$N: \R^{n+1} \to \sph^n$ as
    the function that takes a vector~$\vect v \in \R^{n+1}$ to the unit vector in that
    direction:    \[
        N(\vect{v})=
        \begin{cases}
            \frac{\vect{v}}{\norm{\vect{v}}}
            &\text{if } \vect{v} \neq \vect{0}\\
            (1,0,\ldots, 0)
            &\text{if } \vect{v} = \vect{0}.
        \end{cases}
    \]
\end{definition}

The following defines a closed geodesic ball on the surface of a sphere.

\begin{definition}[Closed Spherical Cap]\label{def:spherical-cap}
    Let~$\vect c \in \sph^n$ and~$r \in [0,\pi]$. Then, the \emph{closed
    spherical cap}
    of (geodesic) radius~$r$ centered at~$\vect c$ is
\[B_r(\vect c) := \{\vect s \in \sph^n ~|~ \cos^{-1}(\vect c \cdot \vect s) \leq r\}
            = \{\vect s \in \sph^n ~|~ \vect c \cdot \vect s \leq \cos(r) \}.
    \]
\end{definition}

Geometrically, this is the intersection of a closed linear half
space with~$\S^n$; see \figref{filtration}.

In the following definition, we introduce a function specifying the radius of the cap associated with~$\vect v$ at a height threshold~$t$.
\begin{definition}[Geodesic Radius of Spherical Cap]
    \label{def:radius}
    Given~$\vect{v} \in \R^{n+1}$, the \emph{geodesic radius function}~$r_{\vect{v}}:
    \R \to [0, \pi]$ is defined, for~$t \in \R$, as the geodesic radius of the spherical cap
    obtained by intersecting~$\S^n$ with the half space at and below height~$t$ in
    direction~$N(\vect{v})$:
    \[
        r_{\vect{v}}(t) =
        \begin{cases}
            \cos^{-1} \left(\max\left\{-1, \min\left\{\frac{-t}{\norm{\vect v}},
            1\right\}\right\}\right)
            &\text{if } \vect{v} \neq \vect{0}\\
            0   &\text{if } \vect{v} = \vect{0} \text{ and } t<0\\
            \pi &\text{if } \vect{v} = \vect{0} \text{ and } t \geq 0.
        \end{cases}
    \]
\end{definition}

From the above definitions, we obtain the following lemma:

\begin{figure}
    \centering
    \begin{subfigure}{0.49\textwidth}
        \includegraphics[width=\textwidth]{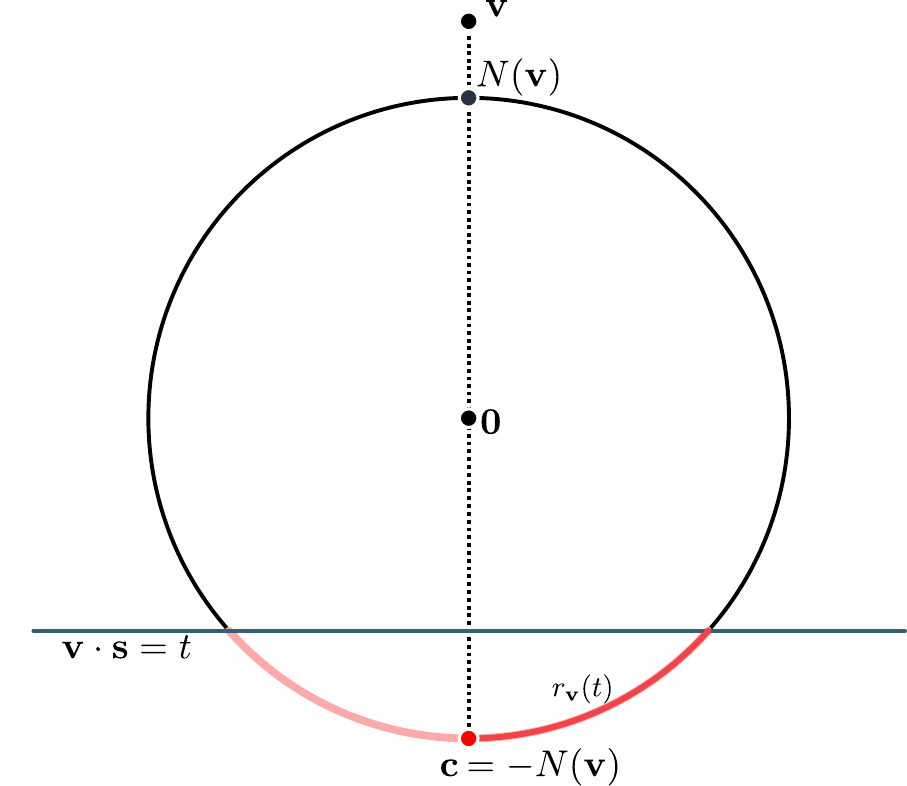}
        \caption{2D spherical cap in~$\S^1$}
        \label{subfig:filtration-first}
    \end{subfigure}
    \hfill
    \begin{subfigure}{0.49\textwidth}
        \includegraphics[width=\textwidth]{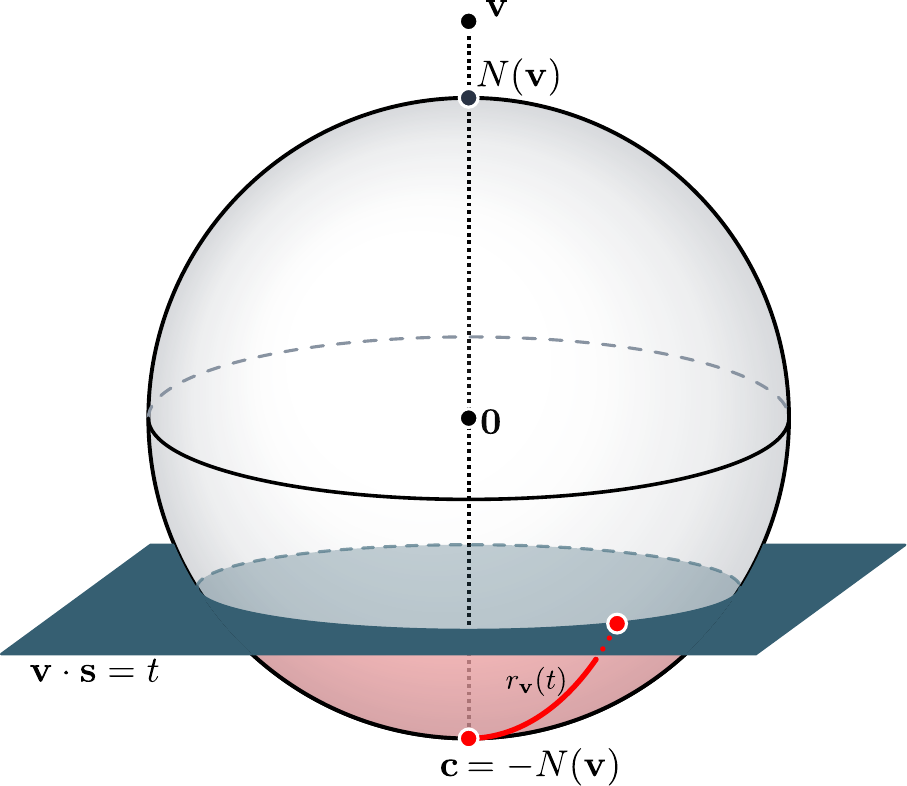}
        \caption{3D spherical cap in~$\S^2$}
        \label{subfig:filtration-second}
    \end{subfigure}
    \caption{Spherical caps (\defref{spherical-cap}). A spherical cap is
    the intersection of~$\S^n$
with the half space defined by the line/plane~$\vect{v}\cdot{s}=t$.}
        \label{fig:filtration}
\end{figure}

\begin{lemma}[Decomposing the Support Using Spherical Caps]\label{lem:intersection}
    Let~$n \geq 1$ and let~$\sigma$ be a geometric simplex in~$\R^{n+1}$.
    Then, for any vertex~$\vect{v} \in \sigma$ and number~$t \in \R$,
        \[
            \mu\big(\suppfcn{\sigma}{t} \cap \vwit{n}{\sigma}{\vect{v}}\big) =
            \mu\big(B_{r_{\vect{v}}(t)} (-N(\vect{v}))
            \cap \vwit{n}{\sigma}{\vect{v}}\big).
        \]
\end{lemma}
\begin{proof}
    Set~$\vect c = -N(\vect v)$. We first observe from the definitions that
    \begin{align*}
        \mu\big(\suppfcn{\sigma}{t} \cap \vwit{n}{\sigma}{\vect{v}}\big) &=
            \mu\Big(\{ \vect{s} \in \vwit{n}{\sigma}{\vect{v}} \mid \vect{v} \cdot
            \vect{s} \leq t \}\Big).
    \end{align*}

    We consider four cases. First, suppose that~$\vect{v} \neq \vect{0}$ and that~$t \in [-\norm{\vect v}, \norm{\vect v}]$.
    Then, letting~$d_{\sph^n}$ denote the geodesic distance on~$\sph^n$,
        \begin{align*}
        \mu\big(\suppfcn{\sigma}{t} \cap \vwit{n}{\sigma}{\vect{v}}\big)
        &= \mu\Big(\Big\{ \vect{s} \in \vwit{n}{\sigma}{\vect{v}} \mid \norm{\vect{v}}
            \cos \Big(d_{\sph^n} (-\vect c, \vect s) \Big) \leq t
            \Big\}\Big)\\
        &= \mu\Big(\left\{ \vect{s} \in \vwit{n}{\sigma}{\vect{v}} \mid d_{\sph^n}
            (-\vect c, \vect s) \geq \cos^{-1} (t / \norm{\vect{v}})
            \right\}\Big)\\
        &= \mu\Big(\left\{ \vect{s} \in \vwit{n}{\sigma}{\vect{v}} \mid d_{\sph^n}
            (\vect c, \vect s) \leq \pi -
            \cos^{-1} (t / \norm{\vect{v}}) \right\}\Big)\\
        &= \mu\Big(\left\{ \vect{s} \in \vwit{n}{\sigma}{\vect{v}} \mid d_{\sph^n}
        (\vect c, \vect s) \leq \cos^{-1} (-t / \norm{\vect{v}}) \right\}\Big)\\
        &= \mu\big(B_{r_{\vect{v}}(t)}(\vect c)
            \cap \vwit{n}{\sigma}{\vect{v}}\big).
    \end{align*}

    Then, suppose that~$t < - \norm{\vect{v}}$. Then,~$r_{\vect{v}}(t) = 0$ and
    \begin{align*}
        \mu\big(\suppfcn{\sigma}{t} \cap \vwit{n}{\sigma}{\vect{v}}\big)
        &=\mu\Big(\{ \vect{s} \in \vwit{n}{\sigma}{\vect{v}}
        \mid \vect{v} \cdot \vect{s} \leq t\}\Big)\\
        &= \mu(\emptyset) = 0 \\
        &= \mu\big(B_{r_{\vect{v}}(t)}(\vect c) \cap \vwit{n}{\sigma}{\vect{v}}\big) \quad
        \text{as } \mu(B_0(\vect c)) = \mu\big(\{\vect c\}\big) = 0.
    \end{align*}

    Now, suppose that~$t > \norm{\vect{v}}$. Then,~$r_{\vect{v}}(t) = \pi$ and
    \begin{align*}
        \mu\big(\suppfcn{\sigma}{t} \cap \vwit{n}{\sigma}{\vect{v}}\big)
        &= \mu\Big(\{ \vect{s} \in \vwit{n}{\sigma}{\vect{v}} \mid
        \vect{v} \cdot \vect{s} \leq t\}\Big)\\
        &= \mu\big(\vwit{n}{\sigma}{\vect{v}}\big)\\
        &= \mu\big(B_{r_{\vect{v}}(t)}(\vect c) \cap
        \vwit{n}{\sigma}{\vect{v}}\big).
    \end{align*}

    Finally, suppose that~$\vect{v} = \vect{0}$.
    Then,
    \begin{align*}
        \mu\big(\suppfcn{\sigma}{t} \cap \vwit{n}{\sigma}{\vect{v}}\big)
        &= \mu\Big(\{\vect{s} \in \vwit{n}{\sigma}{\vect{v}} \mid \vect{v} \cdot \vect{s}
        \leq t\}\Big)\\
        &= \mu\Big(\{\vect{s} \in \vwit{n}{\sigma}{\vect{v}} \mid 0 \leq t\}\Big)\\
        &=
        \begin{cases}
            \mu\big(\vwit{n}{\sigma}{\vect{v}}\big)     &\text{if } t \geq 0\\
            0       &\text{if } t<0
        \end{cases} \\
        &= \mu\big(B_{r_{\vect{v}}(t)}(\vect c)
            \cap \vwit{n}{\sigma}{\vect{v}}\big).\qedhere
    \end{align*}
\end{proof}

We are now ready to prove our main theorem for computing the DIWECT.

\begin{theorem}[Main Theorem]\label{thm:diwect-solution}
    Let~$\K = (K, \omega)$ be a weighted simplicial complex immersed in~$\R^{n+1}$. Set~$\vect c = -N(\vect v)$.
    Then, \[
        \diwect_\K(t)
        =  \sum_{\sigma \in K} (-1)^{\dim \sigma} \omega(\sigma)
            \sum_{\vect{v} \in \sigma} \mu \Big( B_{r_{\vect{v}}(t)}
        (\vect c) \cap \vwit{n}{\sigma}{\vect{v}} \Big). \]
\end{theorem}

\begin{proof}
    The first equality follows from \eqsref{diwectint}{ind-fn-integral},
    so it suffices to show, for each~$\sigma \in K$, the per-simplex identity below.
    By \lemref{partition}, the sets~$\{\vwit{n}{\sigma}{\vect{v}}\}_{\vect{v} \in \sigma}$ form a partition of~$\sph^n$ modulo a set of measure zero. Thus,
    \begin{align*}
        \mu \big( \suppfcn{\sigma}{t} \big) &= \sum_{\vect{v} \in \sigma}
        \mu \big( \suppfcn{\sigma}{t} \cap \vwit{n}{\sigma}{\vect{v}} \big)\\
        &= \sum_{\vect{v} \in \sigma} \mu \Big( B_{r_{\vect{v}}(t)}
        (\vect c) \cap \vwit{n}{\sigma}{\vect{v}} \Big)
        &\text{by \lemref{intersection}}.
        \hspace{20em}\qedhere
    \end{align*}
\end{proof}

In the following two subsections, we show that this admits an
closed-form solution for simplicial complexes in~$\R^2$ or~$\R^3$.

\subsection{Computing the DIWECT in Two Dimensions}

Let~$\K = (K, \omega)$ be a weighted simplicial complex immersed in~$\R^2$.
Here, we use \thmref{diwect-solution} to provide a closed-form solution for
\eqref{diwectint} with~$n = 1$.
We proceed by constructing formulas~for
\begin{equation}
    \int_{\vect{s} \in \sph^1} \ifn_\sigma (\vect{s}, t) \dif \vect{s}
    = \mu \big( \suppfcn{\sigma}{t} \big)
\end{equation}
for the three cases where~$\sigma$ is a zero-simplex, a one-simplex, or a
two-simplex.

Let~$\vect{v} \in K$ be a zero-simplex. By \lemref{partition},~$\vect v$ only
has a single vertex witness region, and~$\vwit{1}{\vect v}{\vect v} = \sph^1$.
Therefore, by \thmref{diwect-solution},
\[
    \mu \big(\suppfcn{\vect{v}}{t} \big) = \mu \Big( B_{r_\vect{v}(t)}
    (\vect c) \Big) = 2 r_{\vect{v}}(t).
\]

If~$\sigma \in K$ is a one-simplex or a two-simplex then by
\thmref{diwect-solution},
\[
    \mu \big(\suppfcn{\sigma}{t} \big) = \sum_{\vect{v} \in \sigma}
    \mu \Big( B_{r_{\vect{v}}(t)} (\vect c)
    \cap \vwit{1}{\sigma}{\vect{v}} \Big).
\]
By \defref{sph-sigma}, each~$\vwit{1}{\sigma}{\vect{v}}$ is an intersection
of semicircles.
To represent these sets, we introduce the following notation.

\begin{definition}
    For~$\vect{u_0}, \vect{u_1} \in \sph^1$, let~$[\vect{u_0}, \vect{u_1}]$ denote the counter-clockwise arc
    from~$\vect{u_0}$ to~$\vect{u_1}$ in~$\S^1$.
\end{definition}

If~$\sigma = \{\vect{v_0}, \vect{v_1}\}$ is a one-simplex, then it follows
immediately from \defref{sph-sigma} that
\begin{equation} \label{eq:one-simplex-2d}
    \closedvwit{1}{\sigma}{\vect{v_0}} =
    \big[ (\vect{v_1} - \vect{v_0})^\perp, (\vect{v_0} - \vect{v_1})^\perp \big]
\end{equation}
where~$\vect{v}^\perp$ is the vector~$\vect{v}$ rotated by 90
degrees counter-clockwise.
Likewise, if~$\sigma = \{\vect{v_0}, \vect{v_1}, \vect{v_2}\}$ is a two simplex
with vertices listed in counter-clockwise orientation, then
\begin{equation} \label{eq:two-simplex-2d}
    \closedvwit{1}{\sigma}{\vect{v_0}} =
    \big[ (\vect{v_2} - \vect{v_0})^\perp, (\vect{v_0} - \vect{v_1})^\perp \big].
\end{equation}

Next, we prove a general formula for the measure of a closed ball intersected
with an interval~$[\vect{u_0}, \vect{u_1}] \subseteq \sph^1$.

\begin{definition}
    For~$\vect{u} = (u_0, u_1)$ and~$\vect{v} = (v_0, v_1)$ in~$\sph^1$, the
    wedge product of~$\vect{u}$ and~$\vect{v}$~is
    \[
        \vect{u} \wedge \vect{v} = u_0 v_1 - u_1 v_0.
    \]
    The signed angle function~$\delta: \sph^1 \times \sph^1 \to (-\pi, \pi]$ is defined as
    \[
        \delta ( \vect{u}, \vect{v} ) = \atantwo ( \vect{u} \wedge \vect{v},
        \vect{u} \cdot \vect{v} ).
    \]
\end{definition}

For any~$\vect{u}, \vect{v} \in \sph^1$, the signed angle~$\delta(\vect{u},
\vect{v})$ is positive if the shortest path from~$\vect{u}$ to~$\vect{v}$ is counter-clockwise, otherwise it is negative and non-unique.
In any case,~$|\delta(\vect{u}, \vect{v})| = d_{\sph^1}(\vect{u}, \vect{v})$.

\begin{definition}
    Let~$\sgn: \R \to \R$ be the modified sign function
    \[
        \sgn(x) =
        \begin{cases}
            1   &\text{if } x \geq 0\\
            -1  &\text{if } x < 0.
        \end{cases}
    \]
\end{definition}

\begin{lemma} \label{lem:2d-intersection-measure}
    Let~$\vect{u_0}, \vect{u_1} \in \sph^1$ with~$\delta(\vect{u_0}, \vect{u_1}) \geq 0$.
    For any~$\vect{s} \in \sph^1$, let~$\alpha = \delta(\vect{u_0}, \vect{s})$ and~$\beta = \delta(\vect{s}, \vect{u_1})$.
    Then,
    \begin{equation*}
        \mu \big( B_r (\vect{s}) \cap [\vect{u_0}, \vect{u_1}] \big) =
        \sgn(\alpha) \min\{ r, |\alpha| \}
        + \sgn(\beta) \min \{ r, |\beta| \}
        + \frac{r}{2}(1 - \sgn(\alpha))(1-\sgn(\beta))
    \end{equation*}
    for all~$r \in [0, \pi]$.
\end{lemma}
\begin{proof}
    The condition that~$\delta(\vect{u_0}, \vect{u_1}) \geq 0$ is equivalent to~$[\vect{u_0}, \vect{u_1}]$ being contained inside a semicircle.
    This ensures that~$\vect{s} \in [\vect{u_0}, \vect{u_1}]$ if and only if~$\alpha, \beta \geq 0$.
    It follows that if~$\alpha, \beta \geq 0$,~then
    \begin{align*}
        \mu \big( B_r (\vect{s}) \cap [\vect{u_0}, \vect{s}] \big)
        &= \min\{ r, \alpha \}\\
        \mu \big( B_r (\vect{s}) \cap [\vect{s}, \vect{u_1}] \big)
        &= \min\{ r, \beta \}.
    \end{align*}
    Note that the term
    \[
        \frac{r}{2} \big( 1 - \sgn(\alpha) \big) \big( 1 - \sgn(\beta) \big)
    \]
    is~$0$ unless~$\sgn(\alpha) = \sgn(\beta) = -1$ in which case it has value~$2r$.
    We now proceed by cases.

    \textbf{Case 1:} Suppose~$\sgn(\alpha) = \sgn(\beta) = 1$.
    Then,~$s \in [\vect{u_0}, \vect{u_1}]$ and so
    \[
        [\vect{u_0}, \vect{u_1}] = [\vect{u_0}, \vect{s}] \cup [\vect{s}, \vect{u_1}].
    \]
    Therefore,
    \begin{align*}
        \mu \big( B_r (\vect{s}) \cap [\vect{u_0}, \vect{u_1}] \big)
        &= \mu \big( B_r (\vect{s}) \cap [\vect{u_0}, \vect{s}] \big)
        + \mu \big( B_r (\vect{s}) \cap [\vect{s}, \vect{u_1}] \big)\\
        &= \min \{ r, \alpha \} + \min \{ r, \beta \}\\
        &= \sgn(\alpha) \min \{ r, |\alpha| \} + \sgn(\beta) \min \{ r, |\beta| \}.
    \end{align*}

    \textbf{Case 2:} Suppose~$\sgn(\alpha) = -1$ and~$\sgn(\beta) = 1$.
    Then,~$[\vect{u_0}, \vect{u_1}] \eqae [\vect{s}, \vect{u_1}] \setminus
    [\vect{s}, \vect{u_0}]$ and~$[\vect{s}, \vect{u_0}] \subseteq [\vect{s},
    \vect{u_1}]$, so
    \begin{align*}
        \mu \big( B_r (\vect{s}) \cap [\vect{u_0}, \vect{u_1}] \big)
        &= \mu \big( B_r (\vect{s}) \cap [\vect{s}, \vect{u_1}] \big)
        - \mu \big( B_r (\vect{s}) \cap [\vect{s}, \vect{u_0}] \big)\\
        &= \min\{r, \beta\} - \min\{r, |\alpha|\}\\
        &= \sgn(\alpha) \min \{ r, |\alpha| \}
        + \sgn(\beta) \min \{ r, |\beta| \}.
    \end{align*}

    \textbf{Case 3:} The case where~$\sgn(\alpha) = 1$ and~$\sgn(\beta) = -1$
    follows from the same argument as the second case.

    \textbf{Case 4:} Suppose that~$\sgn(\alpha) = \sgn (\beta) = -1$, so~$\alpha, \beta \in (-\pi, 0)$.
    Then,
    \begin{align*}
        \delta(\vect{u_1}, \vect{s}) &= |\beta|\\
        \delta(\vect{s}, \vect{u_0}) &= |\alpha|.
    \end{align*}
    and both lie in~$(0, \pi)$.
    Thus, the counter-clockwise arcs~$[\vect{u_1}, \vect{s}]$ and~$[\vect{s}, \vect{u_0}]$
    have lengths~$|\beta|$ and~$|\alpha|$ (respectively).
    Traveling counter-clockwise from~$\vect{u_0}$ to~$\vect{u_1}$, to~$\vect{s}$,
    and back to~$\vect{u_0}$ covers a total length of~$2\pi$.
    Therefore,~$\vect{s} \in [\vect{u_1}, \vect{u_0}]$ and we have~$[\vect{u_1}, \vect{u_0}] = [\vect{u_1}, \vect{s}] \cup [\vect{s}, \vect{u_0}]$.
    Observe that this only occurs when~$\vect{s}$ lies in the arc from~$-\vect{u_0}$ to~$-\vect{u_1}$, but we only require~$\vect{s} \in
    [\vect{u_1}, \vect{u_0}]$.~Therefore,
    \begin{align*}
        \mu \big( B_r(\vect{s}) \cap [\vect{u_1}, \vect{u_0}] \big)
        &= \mu \big( B_r(\vect{s}) \cap [\vect{u_1}, \vect{s}] \big)
        + \mu \big( B_r(\vect{s}) \cap [\vect{s}, \vect{u_0}] \big)\\
        &= \min\{r, |\beta|\} + \min\{r, |\alpha|\}.
    \end{align*}
    Now, because~$[\vect{u_0}, \vect{u_1}] \eqae \sph^1 \setminus [\vect{u_1},
    \vect{u_0}]$,
    \begin{align*}
        \mu \big( B_r(\vect{s}) \cap [\vect{u_0}, \vect{u_1}] \big)
        &= \mu \big( B_r(\vect{s}) \cap \sph^1 \big)
        - \mu \big( B_r(\vect{s}) \cap [\vect{u_1}, \vect{u_0}] \big)\\
        &= 2r - \min\{r, |\beta|\} - \min\{r, |\alpha|\}\\
        &= \sgn(\alpha) \min \{ r, |\alpha| \} + \sgn(\beta) \min \{ r, |\beta|
        \} + 2r. \qedhere
    \end{align*}
\end{proof}

Now, combining \eqsref{one-simplex-2d}{two-simplex-2d} with
\lemref{2d-intersection-measure} and \thmref{diwect-solution}
provides a formula for~$\mu(\suppfcn{\sigma}{t})$ for each~$\sigma \in K$.
By \eqref{diwectint}, this induces a formula for~$\diwect_\K$ for any
weighted simplicial complex~$\K$ immersed in~$\R^2$.
Our implementation of the DIWECT in~$\R^2$ is a straight-forward application of
this formula.

\subsection{Computing the DIWECT in Three Dimensions}

Let~$\K = (K, \omega)$ be a weighted simplicial complex immersed in~$\R^3$.
As such,~$\supp(\sigma, t) \subseteq \sph^2$, and~$\mu$ denotes surface area on the unit
sphere. \thmref{diwect-solution} states~$\mu(\supp(\sigma, t))$ as a sum of
areas obtained by intersecting a spherical cap with vertex witness regions~$\vwit{2}{\sigma}{\vect v}$.
As we build up to the formula for computing the DIWECT in 3D, we
first define the contribution of a single vertex to the surface area of a region
on a sphere.
\begin{definition}[Vertex Contribution]
    For a geometric simplex~$\sigma$ in~$\R^3$ and a vertex~$\vect v$ in~$\sigma$, define~$F_{\sigma, \vect v}(t)$ to be
    the area of the subset of the vertex
    witness region consisting of directions~$s \in \sph^2$ such that~$\sigma$
    has height at most~$t$:
    \[
        F_{\sigma, \vect v}(t) :=
             \mu(\supp(\sigma, t)\cap
            \vwit{2}{\sigma}{\vect v}).
    \]
\end{definition}
\defref{sph-sigma} tells us that~$h_\sigma(\vect s) = h_{\vect
v}(\vect s) = \vect v \cdot \vect s$. Consequently, \[F_{\sigma, \vect v}(t) = \mu\left(\{\vect s \in \vwit{2}{\sigma}{\vect v} \mid \vect v
   \cdot \vect s \leq t \}\right). \]
By \lemref{partition}, the vertex witness regions partition~$\sph^2$ modulo
   sets of measure zero. Thus,

   \[\mu(\supp(\sigma, t)) = \sum_{\vect v \in \sigma}F_{\sigma, \vect v}(t)\]
   and the DIWECT formula becomes

   \[\diwect_\K(t) = \sum_{\sigma \in K }(-1)^{\dim \sigma}\omega(\sigma)\sum_{\vect v
       \in \sigma}F_{\sigma, \vect v}(t).\]

\begin{figure}
    \centering
    \begin{subfigure}[t]{0.48\textwidth}
        \includegraphics[width=\linewidth]{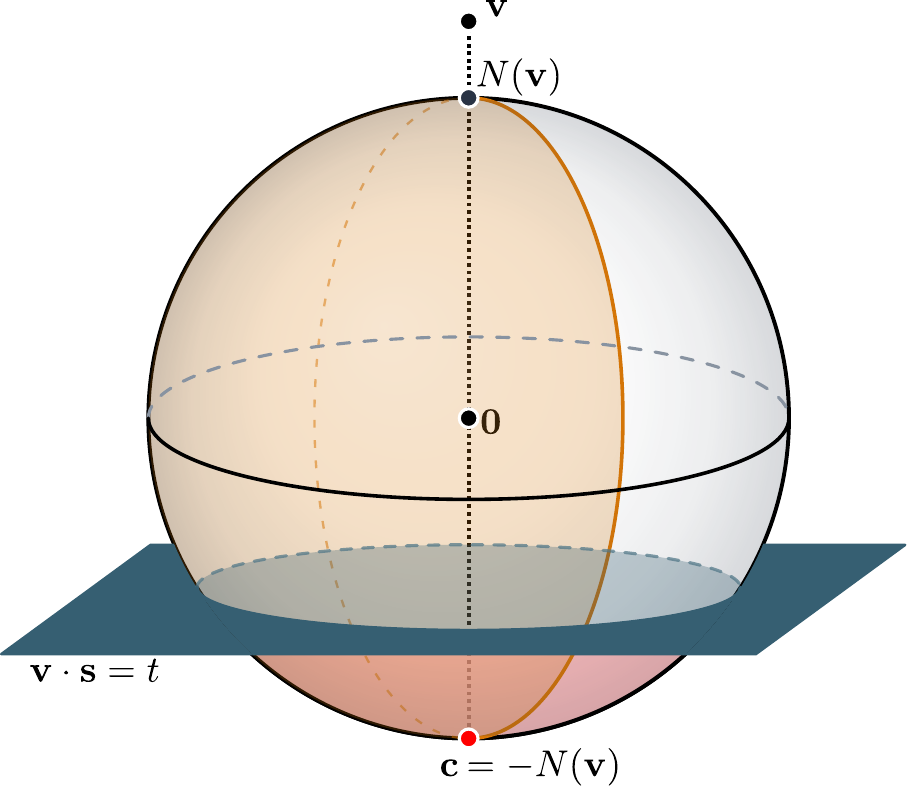}
        \caption{One hemisphere intersecting a spherical cap.}
        \label{subfig:spherical-region-orange}
    \end{subfigure}
    \hfill
    \begin{subfigure}[t]{0.48\textwidth}
        \includegraphics[width=\linewidth]{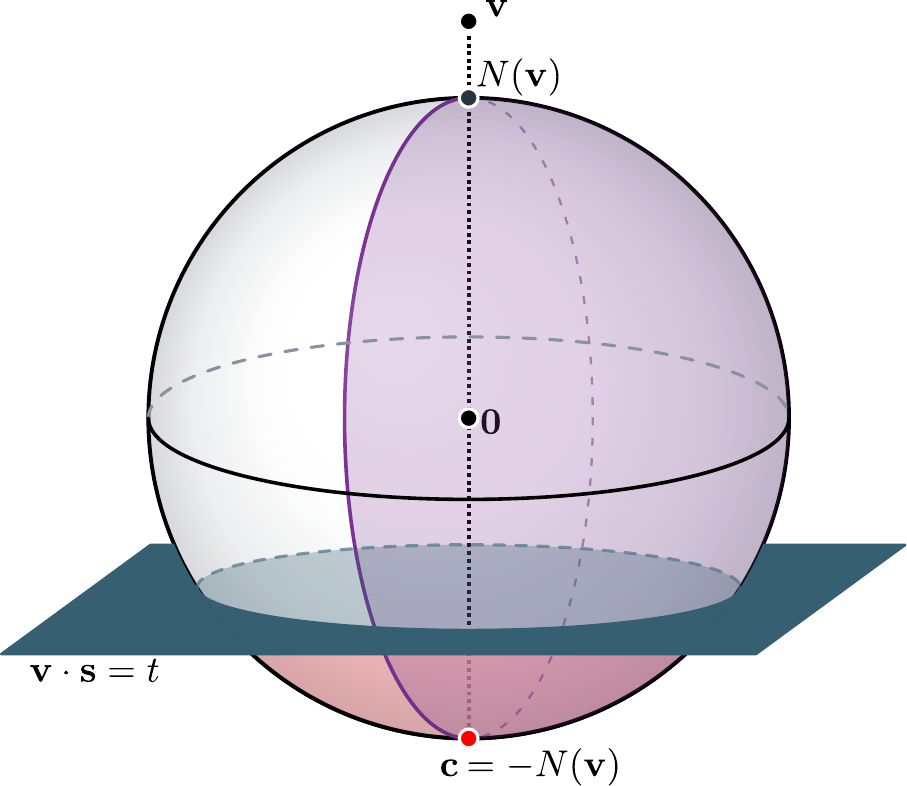}
        \caption{A different hemisphere.}
        \label{subfig:spherical-region-purple}
    \end{subfigure}

    \medskip
    \begin{subfigure}[t]{0.48\textwidth}
        \includegraphics[width=\linewidth]{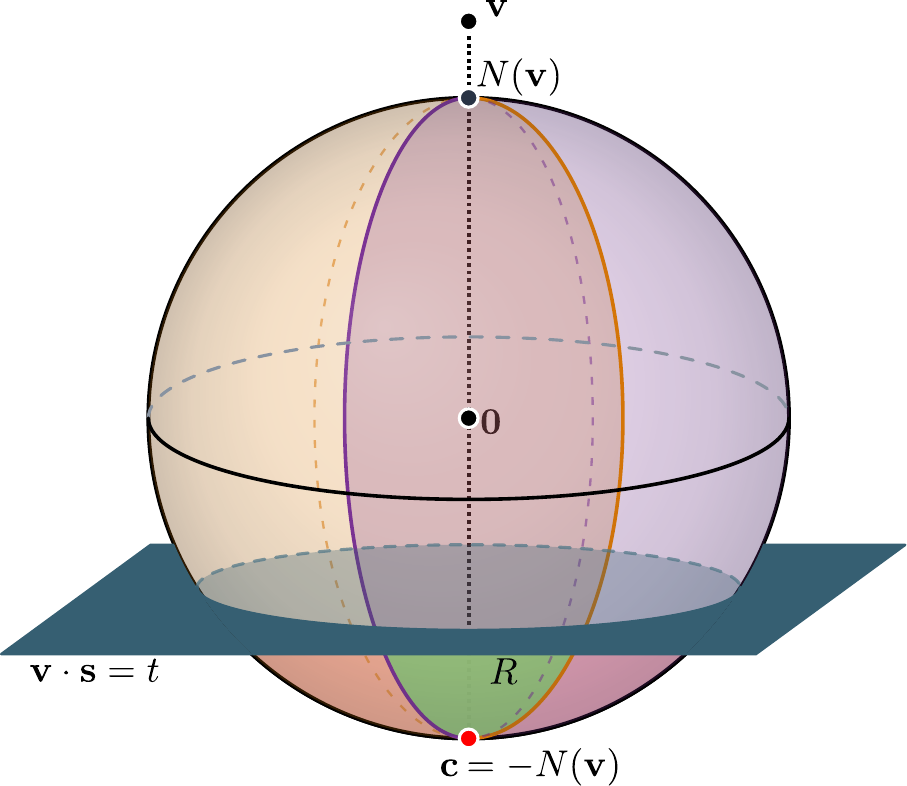}
        \caption{All hemispheres together.}
        \label{subfig:spherical-region-intersection}
    \end{subfigure}
    \caption{A possible spherical region for a two-simplex in 3D. This region~$R$ (in green) is the intersection of the spherical cap with the
    hemispheres induced by the vertex witness regions.}
        \label{fig:spherical-region}
\end{figure}

\begin{figure}
    \centering
    \begin{subfigure}[t]{0.48\textwidth}
        \includegraphics[width=\linewidth]{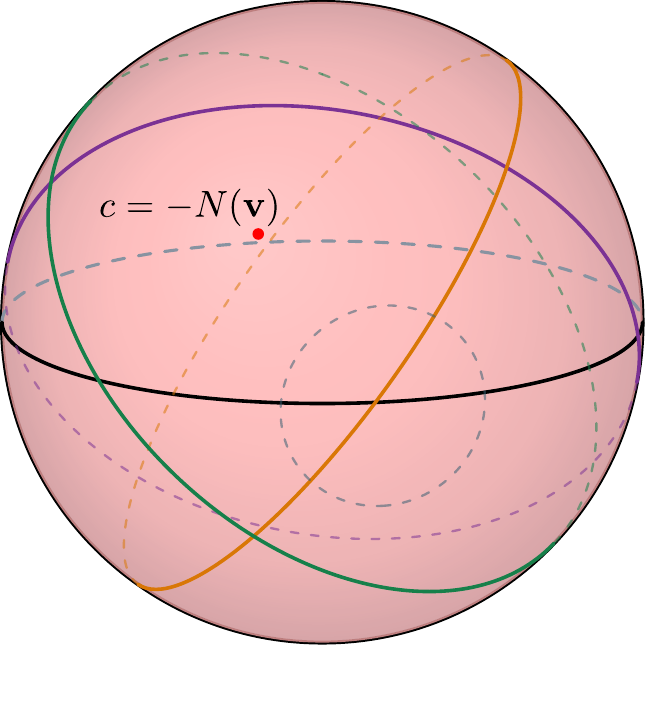}
        \caption{The ``front'' of the unit sphere.}
        \label{subfig:spherical-region-front}
    \end{subfigure}
    \hfill
    \begin{subfigure}[t]{0.48\textwidth}
        \includegraphics[width=\linewidth]{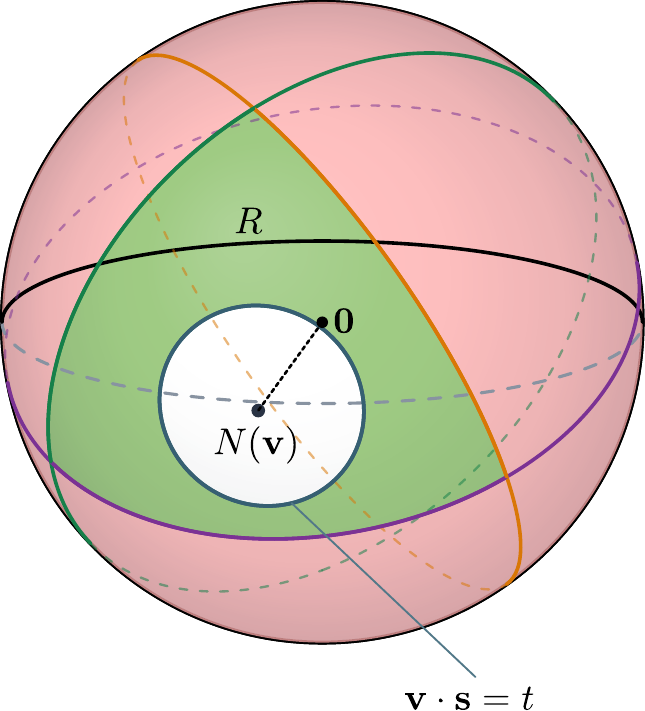}
        \caption{$R$ is a spherical triangle with a circle cut out.}
        \label{subfig:spherical-region-hole}
    \end{subfigure}

    \medskip
    \begin{subfigure}[t]{0.48\textwidth}
        \includegraphics[width=\linewidth]{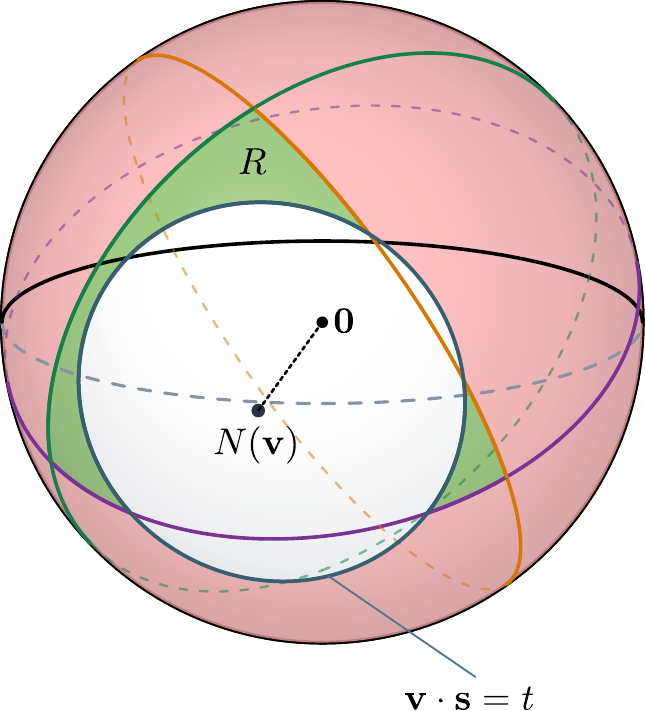}
        \caption{$R$ has two connected components.}
        \label{subfig:spherical-region-two-components}
    \end{subfigure}
    \hfill
    \begin{subfigure}[t]{0.48\textwidth}
        \includegraphics[width=\linewidth]{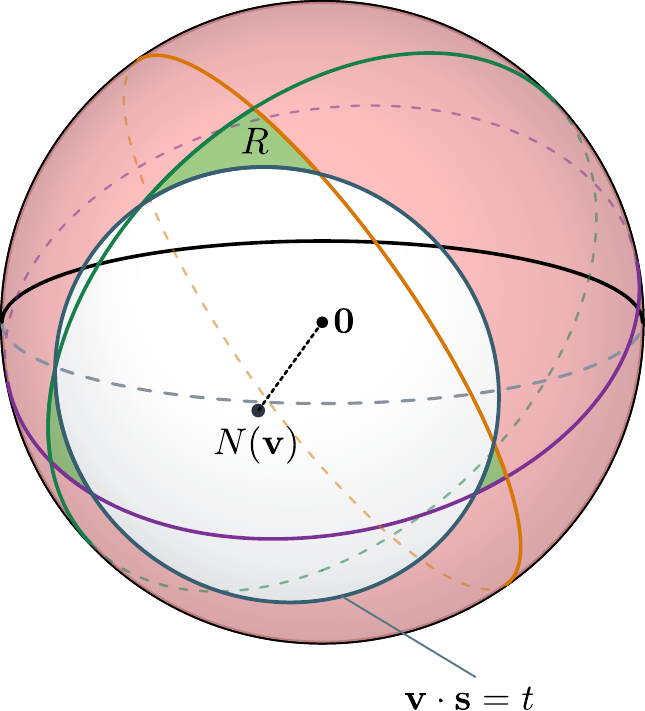}
        \caption{$R$ has three connected components.}
        \label{subfig:spherical-region-three-components}
    \end{subfigure}
    \caption{Pathological examples of what region~$R$ can look like for a
        three-simplex in 3D.
        The green, orange, and purple equators demonstrate hemisphere boundaries.
        \subfigrefrel{spherical-region-front} shows the side of the sphere where
        the filtration begins at~$c$.
        \subfigrefrel{spherical-region-hole}--\subfigrefrel{spherical-region-three-components}
        show the opposite side at different values of~$t$. }
    \label{fig:spherical-region-examples}
\end{figure}

   Let~$R = B_{r_{\vect v}(t)}(\vect c) \cap
   \closedvwit{2}{\sigma}{\vect v}$ be a region on the surface of~$\sph^2$. By \lemref{intersection},~$F_{\sigma, \vect v}(t) =
   \mu(R)$. Thus, computing the contribution of~$v$ to the DIWECT reduces to
   finding the area of~$R$. \figref{spherical-region} demonstrates a simple example of what~$R$ can look like for a~$2$-simplex.
   In \figref{spherical-region-examples}, we visualize three complicated examples of a region~$R$ induced by a~$3$-simplex.
   Note that these are just a few examples of many. As we can see,~$\partial
   R$ consists of a collection of circular arcs. We orient
   each boundary arc such that~$R$ is positively oriented.  Then, we can
   compute~$\mu(R)$ using Stokes' theorem.

\begin{lemma}[Area as boundary integral]
    \label{lem:area-boundary}

       Let~$0 < r_{\vect v}(t) < \pi$. Suppose that~$R$ is a manifold
       with corners and positive area, and~$\partial R$ is oriented positively. Set~$\vect c = -N(\vect v)$. Define the differential one-form
       \[
           \eta_c = \frac{(\vect c \times \vect s) \cdot d\vect s}
           {1+\vect c \cdot \vect s}
       \]
       on~$\sph^2 \setminus \{-\vect c\}$. Then,
       \[\mu(R) = \int_{\partial R} \eta_c.\]

    \end{lemma}

    \begin{proof}
    On~$R$,~$1+\vect c\cdot\vect s\geq 1+\cos r_{\vect v}(t)>0$, so~$\eta_c$ is smooth
    on a neighborhood of~$R$. In positively oriented spherical
    coordinates~$(\theta,\phi)$ with polar axis~$\vect c$,
    \[
        \eta_c=(1-\cos\theta)\,d\phi,
        \qquad
        d\eta_c=\sin\theta\,d\theta\wedge d\phi=dA,
    \]
    where~$dA$ is the sphere's area form. Although these coordinates are
    singular at~$\vect c$, both~$d\eta_c$ and~$dA$ are smooth at this point, so their equality holds
    at~$\vect c$ by continuity. Thus, Stokes' theorem for manifolds
    with corners \cite[Theorem 16.25]{lee2012smooth} gives
    \[
        \mu(R)=\int_R dA=\int_R d\eta_c
        =\int_{\partial R}\eta_c.\qedhere
    \]
\end{proof}

The boundary of~$R$ consists of pieces of boundaries of
the hemispheres and of the spherical cap. We compute different integrals for these
different kinds of~boundaries.

\begin{lemma}[Boundary Decomposition]
    \label{lem:boundary-decomposition}
    Let~$\Gamma_c=\partial B_r(\vect c)\cap R$.
    There are finitely many oriented great-circle arcs~$\gamma_i:\vect a_i\to\vect b_i$ of length less than~$\pi$ such that
    \[
        \int_{\partial R}\eta_c
        =\int_{\Gamma_c}\eta_c+\sum_i\int_{\gamma_i}\eta_c.
    \]
\end{lemma}
\begin{proof}
    Each boundary arc lies on the cap boundary or a hemisphere boundary.
    Assign cap-boundary arcs to~$\Gamma_c$ and count each remaining arc
    once, subdividing it into pieces of length less than~$\pi$.
    Orient these pieces by~$\partial R$. Their interiors are disjoint,
    so additivity of integration gives the result.
\end{proof}

We first examine the pieces of~$\partial R$ contributed by boundaries of
hemispheres, which are great-circle arcs.

\begin{lemma}[Hemisphere-Boundary Contribution]
    \label{lem:great-circle-contribution}

    For~$\vect c \in \sph^2$, let~$\gamma$ be the great-circle arc of length
    less than~$\pi$, oriented from~$\vect a$ to~$\vect b$, with~$-\vect c \notin
    \gamma$. Then

   \[
   \int_\gamma\eta_c=\Omega(\vect c,\vect a,\vect b), \quad \text{where }
       \Omega(\vect c,\vect a,\vect b)
        :=2\atantwo\left(
            \vect c\cdot(\vect a\times\vect b),
            1+\vect c\cdot\vect a+\vect a\cdot\vect b
            +\vect b\cdot\vect c
        \right).
    \]
    \end{lemma}

\begin{proof}
    Let~$T$ be the oriented spherical triangle bounded by
    great-circle arcs~$\vect c\to\vect a\to\vect b\to\vect c$.
    As~$-\vect c\notin\gamma$, this triangle avoids~$-\vect c$.
    The form~$\eta_c$ vanishes along the two sides incident to~$\vect c$.
    Thus, Stokes' theorem identifies~$\int_\gamma\eta_c$ with the signed
    area of~$T$, which is its signed solid angle.
    Van Oosterom and Strackee provide the solid-angle formula
    in~\cite{solidangle}.
    Degenerate triangles follow by continuity.
\end{proof}

\begin{lemma}[Cap-Boundary Contribution]
    \label{lem:cap-boundary-contribution}
    Let~$\Gamma_c$ be as defined in \lemref{boundary-decomposition} and let~$\rho$ be the proportion of the cap-boundary contained in~$R$. Orient~$\Gamma_c$
    by the induced boundary orientation. Then,
    \[
        \int_{\Gamma_c}\eta_c=2\pi(1-\cos r)\rho.
    \]
\end{lemma}

\begin{proof}
        See the formula for~$\rho$ in \lemref{rho}. On~$\Gamma_c$,~$\eta_c=(1-\cos
    r)\,d\phi$, as the polar angle for every point is~$r$.
    The induced orientation has increasing azimuthal angle~$\phi$, and the total
    azimuthal angle is~$2\pi\rho$. Thus
    \[
        \int_{\Gamma_c}\eta_c
        =(1-\cos r)\int_{\Gamma_c}d\phi
        =2\pi(1-\cos r)\rho.\qedhere
    \]
\end{proof}

Combining the contribution of the two types of boundaries gives us the following
formula.

\begin{corollary}[Vertex Contribution in Three Dimensions]
    \label{cor:vertex-contribution-3d}

    \[
        F_{\sigma,\vect v}(t)=\mu(R)
        =\sum_i\Omega(\vect c,\vect a_i,\vect b_i)
        +2\pi(1-\cos r)\rho.
    \]
\end{corollary}

\begin{lemma}[Endpoint and Degenerate Cases]
    \label{lem:area-endpoints}
    For~$r,s\in[0,\pi]$,
    \[
        \mu(B_0(\vect c)\cap\closedvwit{2}{\sigma}{\vect v})=0,\qquad
        \mu(B_\pi(\vect c)\cap\closedvwit{2}{\sigma}{\vect v})
        =\mu(\closedvwit{2}{\sigma}{\vect v}), \text{ and}
    \]
    \[
        \left|\mu(B_r(\vect c)\cap\closedvwit{2}{\sigma}{\vect v})
        -\mu(B_s(\vect c)\cap\closedvwit{2}{\sigma}{\vect v})\right|
        \leq 2\pi|\cos r-\cos s|.
    \]
    Assigning a value of zero to intersections of measure zero, the formula in
    \corref{vertex-contribution-3d} holds for degenerate radii values by
    continuity.
\end{lemma}
\begin{proof}
    The endpoint values follow from~$B_0(\vect c)=\{\vect c\}$ and~$B_\pi(\vect c)=\sph^2$. For~$r\leq s$, nesting of the caps gives
    \begin{align*}
        0&\leq\mu(B_s(\vect c)\cap\closedvwit{2}{\sigma}{\vect v})
        -\mu(B_r(\vect c)\cap\closedvwit{2}{\sigma}{\vect v})\\
        &\leq\mu(B_s(\vect c)\setminus B_r(\vect c))
        =2\pi(\cos r-\cos s).
    \end{align*}
    Interchanging~$r$ and~$s$ follows directly, proving the bound, and
    continuity follows.
    If~$\closedvwit{2}{\sigma}{\vect v}$ has measure zero, every intersection
    has zero area. Otherwise,~$\closedvwit{2}{\sigma}{\vect v}$ is a spherical
    polygon, a hemisphere, or the whole sphere.
    Except at the finitely many radii where the cap boundary meets a
    vertex, is tangent to an edge, or coincides with an edge circle,
    a positive-area intersection is a manifold with corners.
    Apply \corref{vertex-contribution-3d} at these regular radii
    and use continuity at the remaining radii.
\end{proof}

\section{Experiments}
To demonstrate the utility of the DIWECT,
we perform experiments on a synthetic dataset and on a real-world dataset of
barley seed scans.\footnote{
    For reproducibility, the code is available here: \url{https://github.com/compTAG/diwect}
}
The primary goals of these experiments are (i) to demonstrate the advantages
and disadvantages of the rotational invariance inherent to the DIWECT, and
(ii) to compare the computational efficiency (compute time) of
the proposed DIWECT framework with approximations derived from averaging
the~WECT.

\subsection{Datasets and Compute Environment}

The following datasets were used in the experiments:\begin{itemize}
    \item \textbf{Equal Area Shapes 2D} (EAS2D): A synthetic dataset consisting of five
        polygon classes~(bar, L, T, plus, and triangle); see
        \subfigref{exp:1-mechanism}{shapes}. Each shape
        template is
        a~$64 \times 64$ binary image, with~$380$ of those pixels representing
        the shape/foreground (pixel value=$1$), and the remainder being the background
        (value=$0$). Sample data are
        made by rotating a shape template by an angle drawn uniformly at random
        from~$\sph^1$.
        Some experiments also multiply each foreground pixel's weight by random noise.

    \item \textbf{Barley Seeds} (BS3D):~$3{,}121$ X-ray CT scans of individual barley
        seeds at a resolution of~$127\,\mu$m~\cite{amezquita2022measuring},
        collected from~$77$ spikes (seed heads).
        All seeds of a spike share one of~$28$ seed types (the classes), and each
        type has between~$25$ and~$194$ seeds, spread over one or more spikes.
        Each binary scan is
        meshed
        with marching cubes into a triangulated surface (median~$23{,}000$
        triangles),
        and cleaned as in~\cite{amezquita2022measuring},
        including an alignment step where all seeds are aligned using the
        first three principal components.
        See \figref{append:barley-seed} in \appendref{barley-seed}.
\end{itemize}

Experiments on performance are computed on both CPU and GPUs.
CPU performances are measured on
one Intel Xeon core at 2.2GHz with two threads and 12GB of memory. All GPU
computation is done on a single NVIDIA T4 GPU.

\subsection{Structural Properties of DIWECT Feature Vectors}
\label{sec:exp:structure}

Here, we isolate how rotational invariance shapes the
DIWECT's feature space using EAS2D.

\paragraph*{Capturing Shape Information}

\begin{figure}[htbp]
    \centering
    \begin{subfigure}[b]{0.9\textwidth}
        \centering
        \includegraphics[width=\linewidth]{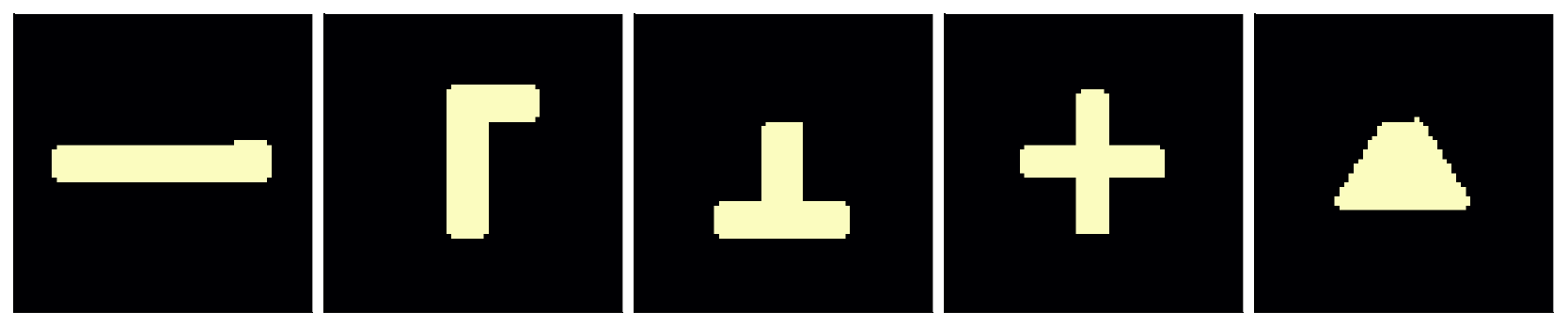}
        \caption{A representative image for each class from the EAS2D dataset.}\label{subfig:exp:1-mechanism-shapes}
    \end{subfigure}

    \vspace{.5em}

    \begin{subfigure}[b]{0.44\textwidth}
        \centering
        \includegraphics[width=\linewidth]{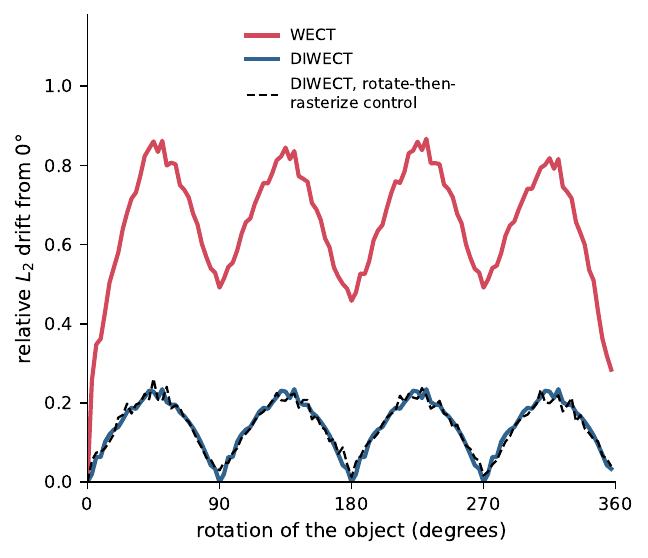}
        \caption{Rotating an image.}\label{subfig:exp:1-mechanism-drift}
    \end{subfigure}
    \hfill
    \begin{subfigure}[b]{0.44\textwidth}
        \centering
        \includegraphics[width=\linewidth]{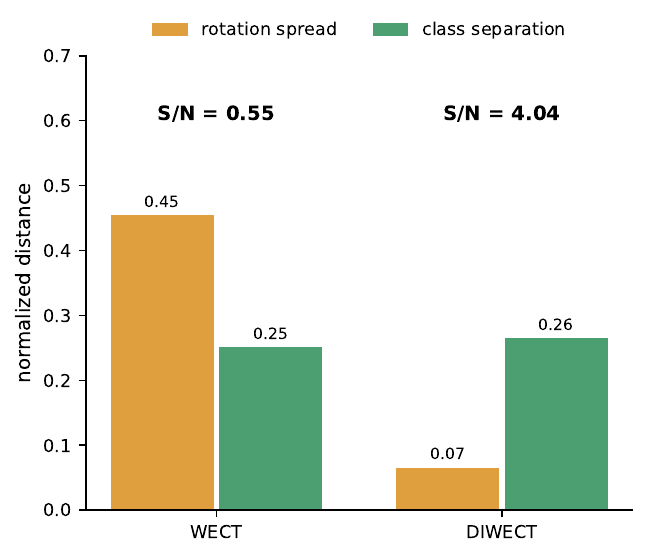}
        \caption{Class spread and separation.}\label{subfig:exp:1-mechanism-signalvsnuisance}
    \end{subfigure}

    \caption{\subfigrefrel{exp:1-mechanism-drift} demonstrates much greater
        spread between rotations of the L shape for the WECT vs the DIWECT. Note
        that the residual~$L_2$ drift between rotations using the DIWECT is due
        to rasterization of the image, rather than the DIWECT itself. In
        \subfigrefrel{exp:1-mechanism-signalvsnuisance} we observe that the
        DIWECT captures similar class separation with a much greater
        signal-to-noise ratio.}
    \label{fig:exp:1-mechanism}
\end{figure}

To investigate effects of rotation in 2D, we rotate the L-shaped
image from EAS2D in steps of~$3^\circ$ and plot each descriptor's~$L_2$ distance from its value under no rotation. Then, for each of the five shapes in the dataset, we
rotate each shape in 36 uniform steps from~$\sph^1$. We measure rotation spread
and class separation within the feature space of the DIWECT and the WECT.
Rotation spread is the mean distance of a rotated sample from its own class
mean, which captures variation due to rotation. Class separation is the mean
distance between class means, capturing shape signal. We compute the signal to
noise (S/N) ratio by dividing class separation by rotation spread, normalized
by mean feature norm. \subfigref{exp:1-mechanism}{drift} and
\subfigrefrel{exp:1-mechanism-signalvsnuisance} demonstrate the results of this
experiment. These results indicate that the DIWECT
does not capture the differences between rotations of the same shape; whereas,
the WECT has large differences between rotations of the same shape. We see
conclude that the DIWECT captures
the same amount of shape information as the WECT, while ignoring~rotation.

\begin{figure}[htbp]
    \centering
    \includegraphics[width=\linewidth]{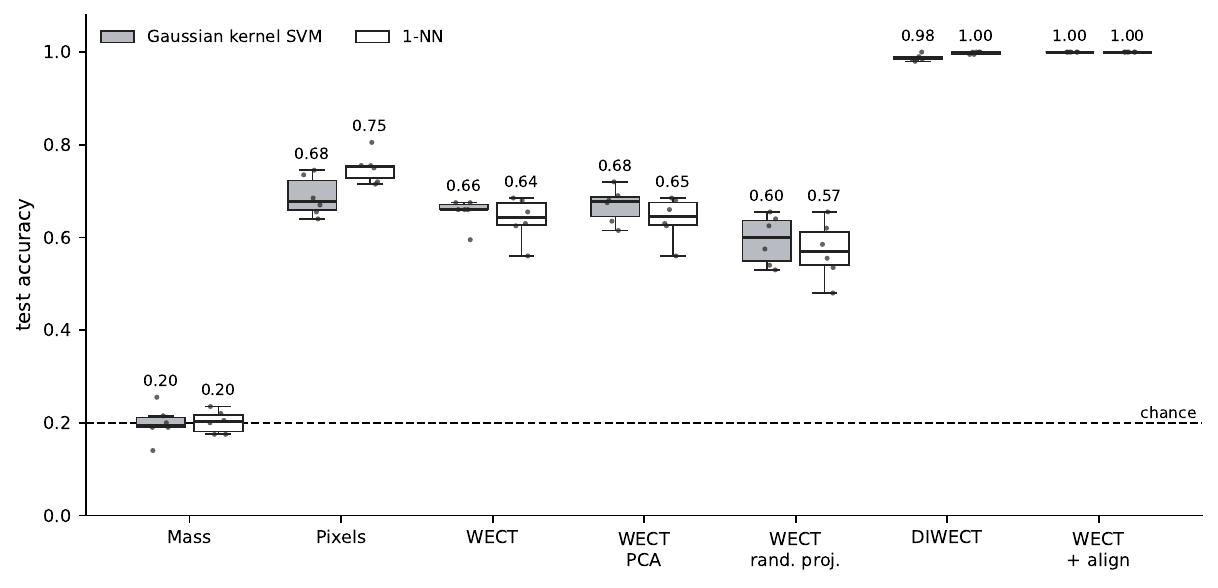}

    \caption{
        For each classifier, the box plot spans the interquartile range with a
        line at the median~(whose value is printed above the box), the whiskers extend to the most
        extreme repetition within~$1.5$ times the interquartile range, and the
        dots are the accuracies of the individual repetitions.
        The feature vectors fall into five categories:
        a control~(Mass),
        feature vectors that are not rotationally invariant (Pixels and WECT),
        dimension controls for the WECT (WECT reduced by PCA and by random projection),
        the DIWECT,
        and the WECT with an oracle that uses the true angle of rotation~(WECT +
        align). Across the board, performance by SVM and 1-NN are similar for
        each distinct feature vector. The DIWECT and aligned WECT greatly
        outperform the other feature~vectors.
            }
    \label{fig:exp:2-dist-classifiers}
\end{figure}

\paragraph*{Rotational Invariance with a Distance-Based Classifier}
Next, we test Gaussian kernel SVM and one-nearest-neighbor for
shape classification on the EAS2D dataset.
Both classifiers compare samples only through distances between their
feature vectors, so they cannot learn to undo a rotation.
These classifiers are trained on several different feature vectors:
mass (number of pixels), raw pixel values, various WECT feature vectors,
and the DIWECT (with 48 heights).
The WECT feature vectors include the following:
the raw WECT (64 directions and 48 heights, resulting in a 3072-dimensional vector),
the WECT reduced by PCA or a random projection (to 48 dimensions),
and the WECT with an alignment oracle (using the true angle of rotation of the
object). Each classifier is run six times independently, each with a newly drawn
random training and test set.
The results are shown in \figref{exp:2-dist-classifiers}. For both classifiers, the control demonstrates that EAS2D
is not separable by pixel volume, as the shapes have equal area.
The feature vectors that are not rotationally invariant
do not accurately learn the class separation.
Furthermore, controlling for dimension (as the WECT results in a much
higher-dimensional feature space) achieves similar or slightly worse results.
However, the distance-based
classifiers can effectively use the feature vectors of the DIWECT to learn
the classification task, as it is rotationally invariant.
Finally, applying a rotation oracle to the WECT enables learning,
demonstrating that rotational noise degrades classification unless
accounted for in the feature vectors.

\paragraph*{Feature Space Geometry}
Finally, we explore the geometry of the feature spaces of the
WECT and DIWECT.
First, we measure within-class distance ratios (without applying any algorithms
or clustering) on each type of feature vector: the raw pixels, the WECT, the WECT with PCA
dimension control, the DIWECT, and the WECT with an alignment oracle.
The within-class distance ratio is the mean squared distance over all pairs of
distinct samples of the same class, divided by the mean squared distance over
all pairs of distinct samples.

\begin{figure}[htbp]
    \centering
    \begin{subfigure}[b]{0.344\textwidth}
        \centering
        \includegraphics[width=\linewidth]{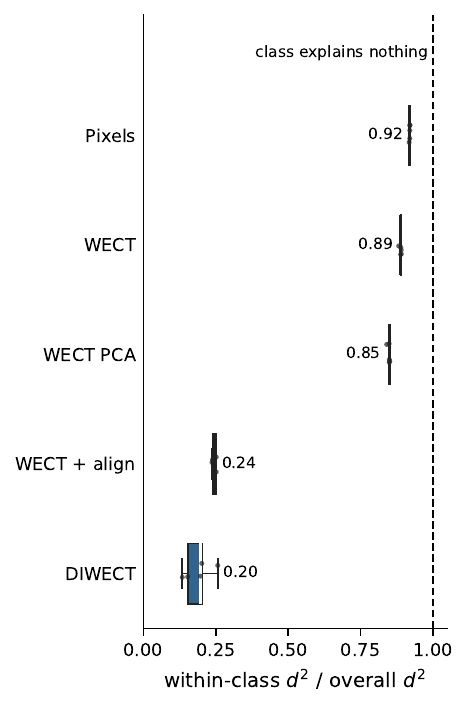}\par
        \vspace{0.132\linewidth}
        \caption{Within-class distance ratios.}\label{subfig:exp:3-feature-geometry-within-class}
    \end{subfigure}
    \hfill
    \begin{subfigure}[b]{0.58\textwidth}
        \centering
        \includegraphics[width=\linewidth]{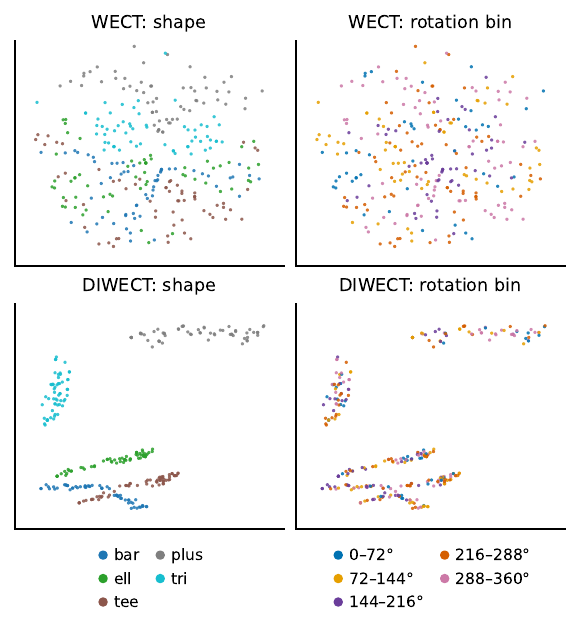}
        \caption{MDS}\label{subfig:exp:3-feature-geometry-mds}
    \end{subfigure}

    \caption{
        \subfigrefrel{exp:3-feature-geometry-within-class} Within-class distance ratios, over five independent repetitions, each with a newly drawn
            random set of rotations.
            A ratio of one indicates that same-shape images are no closer than
            random pairs of images, while zero indicates that they are very
            close.
            The raw pixels, the WECT, and the WECT with PCA
            have ratios near one, while the DIWECT and WECT+align perform much
            better.
            \subfigrefrel{exp:3-feature-geometry-mds}
            An MDS projection of the WECT feature vectors (top row) and the
            DIWECT feature vectors (bottom row).
            The left column colors the projection by shape class, and the right
            column by rotation bin.
            The WECT feature vectors do not clearly separate the shapes,
            while the DIWECT feature vectors group the samples by shape
            (but not rotation).
    }
    \label{fig:exp:3-feature-geometry}
\end{figure}

Next, we project the WECT and DIWECT feature vectors of 300 samples (60
random alignments of each shape) to~$\R^2$ using multidimensional scaling (MDS) on
the Euclidean distances between their feature vectors,
and color the result by shape class and rotation bin.
Results of this experiment are found in \figref{exp:3-feature-geometry}, and
indicate that the feature space of DIWECT and aligned WECT captures the
information about the shape more efficiently, by grouping similar shapes
closely and keeping dissimilar shapes far apart.

\subsection{Analysis of Real-World Barley Seed Data}
\label{sec:exp:barley}

We have seen that the DIWECT is useful for comparing shapes without worrying
about aligning them first. Next, we use the barley seed data, which is already
aligned, to test if the DIWECT can compare against the WECT for pre-aligned data.

\cite{amezquita2022measuring} performs a classification task on barley seed
data using the ECT, and applies an alignment and registration pipeline built
specifically for the seed data before computing the ECT.
In particular, each seed is rotated into the frame of its principal axes
(computed via PCA of its centered voxel coordinates),
and the signs of the axes are fixed by hand-chosen conventions
(e.g., the crease of the seed facing up).
Then, the ECT is sampled in 158 directions concentrated around the crease,
with 16 heights per direction. We reproduce the published ECT values exactly.
Because the seeds are elongated, their principal axes are well
separated: no seed has a ratio above~$0.8$ between consecutive principal
variances. Thus, use of a directional transform is nearly perfect for this
dataset.

\paragraph*{Experimental Setup}
We compute the DIWECT of each seed's surface mesh with 64 heights, normalized by
the seed's radius, and the WECT with the 158 directions and 16 heights
of~\cite{amezquita2022measuring} on the same height grid. We do so in the
following alignments:
\begin{itemize}
    \item the hand-built frame of~\cite{amezquita2022measuring};
    \item the hand-built frame then a rotation of~$\theta \in \{2, 5,
        10, 20, 90\}$ degrees about a random axis
    \item the hand-built frame with the errors that PCA makes when
        principal axes are nearly degenerate, drawn
        independently for every seed: random sign flips of the axes, a swap of
        the width and thickness axes (with probability~$1/2$), or both
    \item a frame chosen without dataset-specific conventions: principal axes
        of the surface, with the first two oriented so that the surface's
        skewness (third moment) along each is positive and the third
        following the right-hand rule (automatic PCA), the
        raw scanner coordinates, or a uniformly random rotation.
\end{itemize}
We also use the 11 \emph{traditional descriptors} of each seed,
published with~\cite{amezquita2022measuring}: measurements of the seed's
length, width, and height, its surface area and volume, the surface area and
volume of its convex hull, and the ratios of the seed's area and volume to
those of its convex hull.
\cite{amezquita2022measuring} classifies the seeds with the ECT combined
with these descriptors, so we also report each topological feature appended to
them, with the traditional descriptors standardized and treated as one more
feature block.
Each feature block is centered and scaled to unit mean squared norm, and the~$28$ seed types are classified with a Gaussian kernel SVM.
We use two splits, each repeated 20 times.
The \emph{seed split} of~\cite{amezquita2022measuring} holds out 25\% of the
seeds of every type at random.
The \emph{spike split} holds out one whole spike for every type with at
least two spikes (which is 26 of the 28 seed types).
Seeds of the same spike were grown together and scanned together in the same
alignment, so the seed split allows a classifier to recognize the spike
rather than the seed type. The spike split removes this shortcut.

\begin{figure}[htbp]
    \centering
    \includegraphics[width=\linewidth]{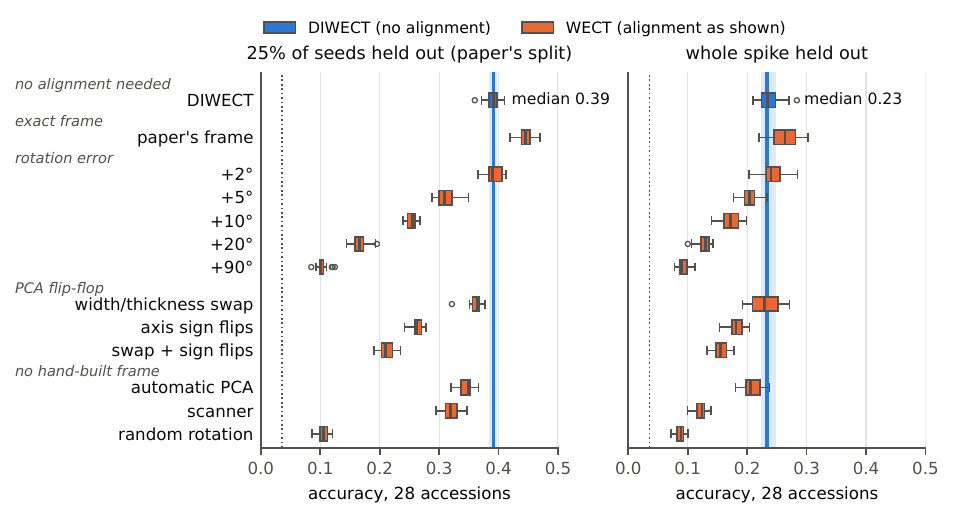}
    \caption{
        Accuracy of classifying 28 barley types over 20 splits with the WECT
        (orange boxes) as the orientation of the seeds degrades, compared with the
        DIWECT (blue box, top row), which does not depend on orientation.
        The DIWECT's median (blue line) and interquartile range (shaded band)
        extend down through every WECT row for comparison.
        Left: 25\% of the seeds of each type held out at random, as
        in~\cite{amezquita2022measuring}.
        Right: one whole spike per type held out.
        Each box spans the interquartile range with a line at the median, the
        whiskers extend to the most extreme split within~$1.5$ times the
        interquartile range, circles are splits beyond the whiskers, and the
        dotted line is chance ($1/28$).
        Only the hand-built frame of~\cite{amezquita2022measuring}, exactly as
        constructed, lets the WECT outperform the DIWECT.
    }
    \label{fig:exp:barley-pose}
\end{figure}

\paragraph*{Results}
\figref{exp:barley-pose} shows the results, and \tabref{exp:barley} in
\appendref{barley-table} lists the mean accuracies in full, together with
the traditional descriptors.
The WECT outperforms the DIWECT only in the hand-built frame, and only by a
small margin (mean accuracy~$0.45$ versus~$0.39$ on the seed split, and~$0.26$ versus~$0.24$ on the spike split).
However, this advantage is disappears under a rotation error of only two degrees.
Furthermore, the sign flips and axis swaps under degenerate principal axes place
the WECT well below the DIWECT ($0.21$--$0.36$ on the seed split), while the
DIWECT remains unchanged.
Without the curated frame, the DIWECT outperforms every WECT alignment, including a
standard automatic PCA frame ($0.35$ on the seed split and~$0.21$ on the spike
split).
Additionally, the scanner frame demonstrates how orientation can also encode
nuisance information:
seeds of a spike are scanned together in the same orientation,
so the WECT in scanner coordinates can recognize the spike rather than the seed type.
In particular, the WECT scores~$0.32$ on the seed split but only~$0.12$ on the spike split (half the accuracy of the DIWECT).

Combined with the traditional descriptors (see \tabref{exp:barley}), the DIWECT is within one standard
deviation of the WECT in the hand-built frame ($0.60$ versus~$0.62$ on the seed
split, and~$0.31$ versus~$0.33$ on the spike split).
In contrast, a WECT in a random or flipped alignment makes the classifier
worse than the traditional descriptors alone.
The DIWECT achieves this with a feature vector of 64 entries per seed (one
per height), compared with~$158 \times 16 = 2{,}528$ entries for the WECT (one
per direction and height, flattened into a single vector before~classification).

The reported accuracies in the above results are lower than the~$0.75$ reported in~\cite{amezquita2022measuring} on the seed split because
our pipeline differs from the published pipeline in three ways aside from alignment.
First, \cite{amezquita2022measuring} computes the ECT on the cubical complex of
the voxels; whereas, we use the surface mesh, which has about twenty times fewer
simplices.
This choice does not appear to affect the accuracy of the DIWECT;
see \appendref{barley-voxels}.
Second, \cite{amezquita2022measuring} bin the ECF between the seed's minimum
and maximum height for each direction;
whereas, we sample every direction on one common grid of heights.
The DIWECT requires a common grid as it averages the EC
over all directions for each height calculated, and we use the same grid for the WECT so
that both transforms are evaluated at the same heights.
Third, \cite{amezquita2022measuring} reduces the ECT to 12 dimensions with
UMAP~\cite{McInnes2018}, fitted without labels on all seeds (including the
test seeds), and trains the SVM on the result; whereas, we train the SVM on the
features directly.
We reproduce the published pipeline~($0.77$ on the seed split), but find that
its accuracy falls to~$0.08$ on the spike split, and still reaches~$0.63$ on
the seed split when every seed is randomly rotated.
Thus, its accuracy mostly reflects which spike a seed came from, rather than
its seed type, and is largely insensitive to the alignment, which is the effect we
wish to measure.
We therefore omit UMAP, so that differences in accuracy reflect the features
themselves.
Our features also lose accuracy on the spike split, but far less (e.g., the
DIWECT falls from~$0.39$ to~$0.24$).

\subsection{Computational Efficiency}\label{sec:exp:efficiency}

As in \cite{bubbles}, the DIWECT may be approximated by averaging the WECT
across a fixed number of directions.
In the following experiments, we compare this baseline with an implementation
of the proposed DIWECT framework in this manuscript.
The approximated DIWECT is computed using \texttt{pyECT}~\cite{SoCGpyECT},
which is a PyTorch-based package that leverages hardware acceleration in
compute environments with a GPU. We use three families of complexes:
\begin{itemize}
    \item Grayscale images ($64^2$ to~$1024^2$ pixels) of a smooth random field on a
        disk, as weighted Freudenthal triangulations
    \item Binary images (the brighter half of the same fields), as cubical complexes
        and as their triangulations
    \item A barley seed from BS3D, as a surface mesh and voxel complex at
        several resolutions.
\end{itemize}
The sizes of these complexes range from~$4300$ to four million cells.
We evaluate the transforms at~$h \in \{17, 65, 257\}$ evenly spaced heights in~$[-M, M]$, where~$M$ is the largest vertex norm.
Because our implementation computes the DIWECT exactly (up to floating-point
error), we treat its output at double precision as the ground truth, and
measure how well the averaged WECT (and our implementation at single
precision) approximates it.
The error of an approximation is defined as its largest absolute deviation
from the ground truth, relative to the largest absolute value of the ground truth.
We normalize by the largest absolute value of the ground truth so that the
error does not depend on the overall scale of the transform~(for example, it is
unchanged when every weight is multiplied by a constant), which makes errors
comparable across complexes; an error of~$10^{-2}$ thus means that no height deviates from the ground truth by more
than~$1\%$ of the largest value of the ground truth.
We use the largest deviation, rather than an~$L_1$ or~$L_2$ distance, so that
the error bounds the worst height rather than an average over heights.

\begin{figure}[htbp]
    \centering
    \includegraphics[width=\textwidth]{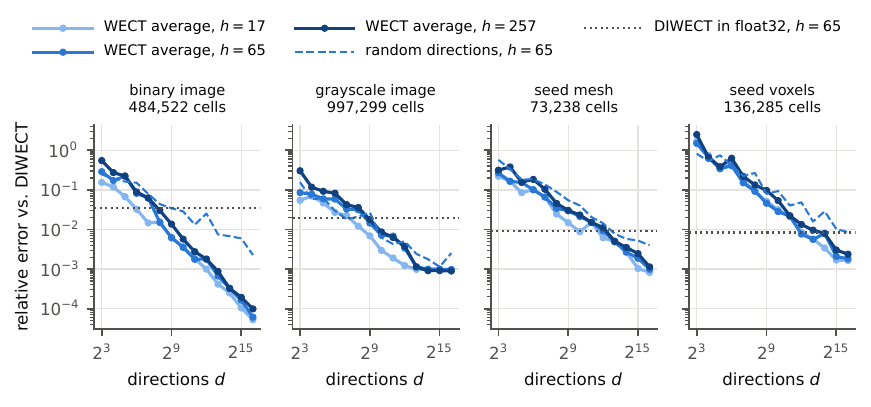}
    \caption{
        Error of the WECT averaged over~$d$ directions, relative to the DIWECT
        computed at double precision,
        for~$h \in \{17, 65, 257\}$ heights (solid) and for random directions (dashed).
        The dotted line is the error (defined in the same way) of the DIWECT
        computed by our implementation in single precision, relative to the
        same implementation in double precision.
    }
    \label{fig:exp:convergence}
\end{figure}

\paragraph*{Convergence}
\figref{exp:convergence} shows that the averaged WECT converges to the DIWECT
as the number of directions~$d$ grows,
for \texttt{pyECT}'s evenly spaced directions (Fibonacci-spiral directions in 3D)
and for i.i.d.\ uniformly chosen directions.
We observe that an error of~$10^{-2}$ requires 256--2{,}048 directions in 2D,
but requires 2{,}048--32{,}768 in 3D.
For the grayscale images, the error levels off near~$10^{-3}$, likely because
\texttt{pyECT} computes in single precision.
For this reason, we also report the DIWECT computed in single precision,
whose error grows with the size of the complex:
below~$1\%$ on complexes of up to about~$1.4 \times 10^5$ cells,~$2\%$--$3.5\%$ on
the~$512^2$ images shown ($0.5$--$1$ million cells), and~$12\%$--$31\%$ on the
largest complexes tested ($1.25$--$4$ million cells, not pictured).

\begin{figure}[htbp]
    \centering
    \includegraphics[width=\textwidth]{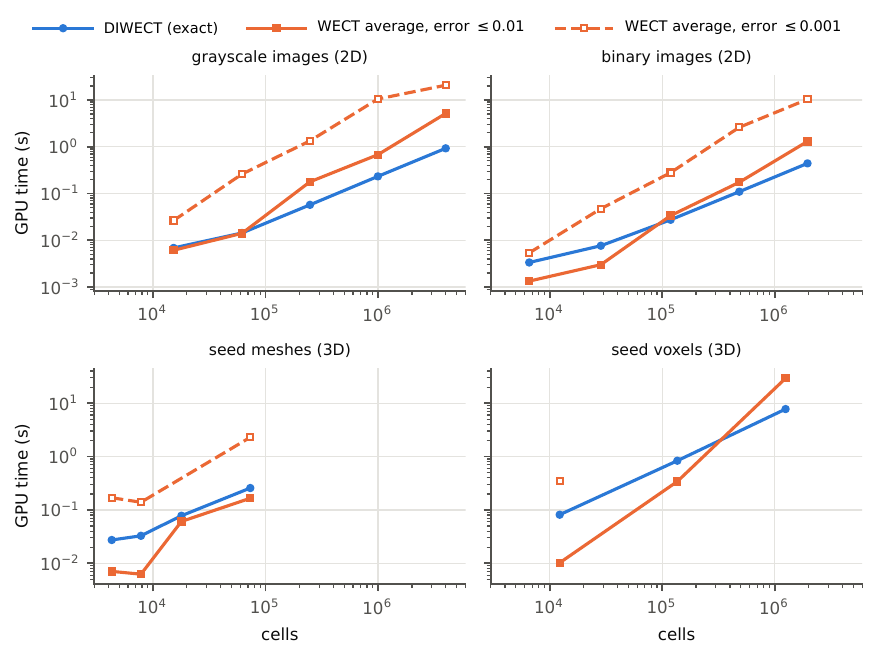}
    \caption{
        Time on an NVIDIA T4 GPU for the DIWECT (blue) and for the WECT (orange)
        averaged over the smallest
        number of directions whose error is at most~$10^{-2}$ (solid) or~$10^{-3}$ (dashed), at~$h = 65$ heights.
        Images and seed voxels are triangulated.
        Missing points mean that error was not reached with up to~$65{,}536$
        directions, the largest number we tried.
    }
    \label{fig:exp:cost}
\end{figure}

\paragraph*{Cost at Matched Accuracy}
\figref{exp:cost} compares the time to compute the DIWECT with the time to
average the WECT over the smallest number of directions that reaches an error
of~$10^{-2}$ or~$10^{-3}$.
For {\it smaller} complexes in 2D,
the averaged WECT with the larger error budget ($10^{-2}$) tends to be equal or
faster than the DIWECT (where both take tens of milliseconds).
However, as the complex size increases, the DIWECT is faster.
The averaged WECT with the smaller error budget~($10^{-3}$) is consistently slower
than the DIWECT at every complex sizes.
For example, on a~$1024^2$ grayscale image (one million cells)
the DIWECT takes 0.23 seconds, while the
averaged WECT takes 0.68 seconds to reach~$10^{-2}$ and 10 seconds to reach~$10^{-3}$.
In 3D, the DIWECT is consistently faster than the averaged WECT with the
smaller error budget ($10^{-3}$) at every size where the averaged WECT reaches that error.
For the averaged WECT with the larger error budget ($10^{-2}$),
the DIWECT tends to be slower for smaller complex sizes.
For larger complex sizes, the averaged WECT ($10^{-2}$) is slightly faster than
the DIWECT for the seed meshes, but the DIWECT is faster for the seed voxels.
The DIWECT's cost grows linearly in the number of cells and in~$h$.
In contrast, the cost of one WECT is almost independent of~$h$, so a finer
height grid does favor the averaged WECT (although it may introduce error).

Finally, we note that both methods bound their memory use.
Our implementation of the DIWECT processes cells in chunks of a fixed number of
(cell, vertex, height) triples, so its peak memory does not grow with the size
of the complex.
Throughout these experiments, the peak GPU memory is roughly 150\,MB for every 2D
complex (including the one with four million cells) and at most 0.8\,GB for the
3D complexes.

\begin{figure}[htbp]
    \centering
    \includegraphics[width=\textwidth]{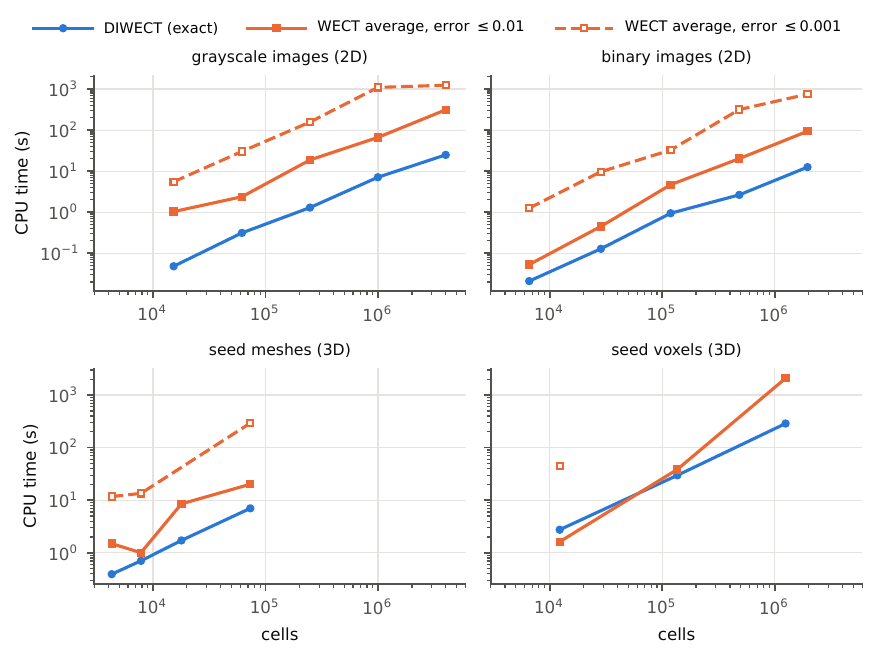}
    \caption{
        As \figref{exp:cost}, but on a two-threaded CPU.
        For each complex and error bound, the number of directions matches the
        GPU experiment.
    }
    \label{fig:exp:cost-cpu}
\end{figure}

\paragraph*{CPU}
We utilize the same parameter choices as in the experiment with the GPU; see \figref{exp:cost-cpu}.
As expected, both methods are slower on the CPU, but the runtime of the averaged WECT
is impacted more.
On complexes above~$1.5 \times 10^4$ cells, the DIWECT is 16--41 times slower
on the CPU than on the GPU, while the averaged WECT is 60--204 times slower.
On the smallest complexes, both slow down less, by factors from 4 and 25, respectively.
As a result, the DIWECT is faster than the averaged WECT with the larger error
budget ($10^{-2}$) on every 2D complex and on every seed mesh, including the
smallest (0.39\,s against 1.5\,s).
The only exception is the smallest triangulated seed voxel complex.
Compared to the averaged WECT with the smaller error budget ($10^{-3}$),
the DIWECT is 16--155 times faster.
For example, on the grayscale image with one million cells, the DIWECT takes
7.1\,s, while the averaged WECT takes 66\,s to reach~$10^{-2}$ and 18 minutes
to reach~$10^{-3}$.
On the full-resolution voxel seed data, the DIWECT takes 4.8 minutes, against
35 minutes for the averaged WECT at~$10^{-2}$.
The DIWECT's peak memory remains bounded on the CPU: at most 0.25\,GB for the 2D
complexes and 0.9\,GB for the triangulated 3D complexes.

\section{Conclusion}

Here, we introduced the DIWECT, which is the average WECT of a
weighted simplicial complex over all filtration directions.
Instead of choosing a fixed sample of directions,
we separate~$\sph^n$ into vertex witness regions for each simplex,
so that computing the DIWECT reduces to measuring the area of region
intersections on the sphere.
In~$\R^2$ and~$\R^3$, these are arc lengths and
spherical areas, which we compute in closed form.
The result is a compact descriptor (a single function of filtration height),
that is invariant to rotation and reflection.
Experimentally, we demonstrate that the DIWECT encodes shape information while
ignoring positional noise.
On the barley seed data, the DIWECT is more accurate in all cases except for the
hand-curated frame of~\cite{amezquita2022measuring}, where the WECT performs
better. However,  the WECT's advantage immediately disappears under a small
rotation error,
while the DIWECT is unaffected by the choice of alignment.
The DIWECT is also faster to compute in most cases than an averaged WECT of
matched accuracy, especially on large complexes and at high accuracy.
Finally, because the centering and integration that define the SECT are linear,
we note that our framework can be easily adapted to compute DETECT~\cite{marsh2024detecting} without needing to sample~directions.

By design, the DIWECT removes the need to provide a canonical positioning or
rotational alignment step before applying the topological descriptor.
However, in cases where a dataset has a reliable framing provided,
the WECT can exploit alignment information while the DIWECT cannot.
In the future, interesting applications include extending the closed-form
implementations to complexes immersed in dimensions higher than three,
as the partition into vertex witness regions holds in any dimension.
Furthermore, a study combining the DIWECT with a few directional WECTs to
recover some alignment information while retaining most of the invariance
appears worthwhile.
Finally, a more thorough study of differentiable relaxations of the DIWECT
(such as those in \cite{roell2023differentiable}) would explore the benefits
of this descriptor in gradient-based learning pipelines.

\section*{Acknowledgments}

{This work is partially funded under National Science Foundation grants 2046730, 2152737, 2337243, and 2346683.
    The authors thank Jack Ruder for his help initially starting this project.
    Also, David Millman was helpful in testing the code for reproducibility.
    Finally, the Computational Topology and Geometry (CompTaG)
lab at Montana State University provided invaluable discussions and feedback
throughout this project.}

\bibliography{references}

\newpage
\appendix
\section{Circular Intervals}

In order to compute the value of~$\rho$ for the 3D DIWECT, we implicitly use
some properties of circular intervals, which are given in this section.

\begin{definition}
    A set~$I \subseteq S^1$ is an \emph{interval} if there exists a halfspace~$H_\vect s (t) \subseteq \R^2$ such that
    \[
        S^1 \cap H_\vect s (t) = I.
    \]
\end{definition}

\begin{definition}
    A \emph{lift} of an interval~$I \subseteq S^1$ is a real interval~$[a,b] \subseteq \R$ with~$a \leq b \leq a + 2 \pi$ such that
    \[
        I = \{ (\cos x, \sin x) \mid x \in [a,b] \}.
    \]
    The \emph{principal lift} of~$I$ is the unique lift~$[a,b]$ such that~$a \in [0, 2\pi]$.
\end{definition}

\begin{lemma} \label{lem:double-circle-intersection}
    Let~$[a_1, b_1]$ and~$[a_2, b_2]$ be principal lifts of circular intervals~$I_1$ and~$I_2$ respectively.
    Then
    \[
        \mu(I_1 \cap I_2) = \sum_{i=-1}^1 \big( \min \{ b_1, b_2 +2\pi i \} - \max \{ a_1, a_2 + 2\pi i\} \big)^+
    \]
    where~$x^+ = \max\{x,0\}$.
\end{lemma}
\begin{proof}
    Within~$[a_1,b_1]$, the lifts of~$I_2$ are the intervals~$[a_2+2\pi i,b_2+2\pi i]$. Only~$i\in\{-1,0,1\}$ can
    overlap~$[a_1,b_1]$ in positive length. These overlaps have
    disjoint interiors and lengths~$(\min\{b_1,b_2+2\pi i\}-\max\{a_1,a_2+2\pi i\})^+$.
    Summing gives the result.
\end{proof}

\begin{lemma} \label{lem:triple-circle-intersection}
    Let~$[a_1, b_1]$,~$[a_2,b_2]$, and~$[a_3, b_3]$ be principal lifts of circular intervals~$I_1$,~$I_2$, and~$I_3$ respectively.
    Then
    \[
        \mu(I_1 \cap I_2 \cap I_3) = \sum_{(i,j) \in \{-1,0,1\}^2} \big( \min \{ b_1, b_2 +2\pi i, b_3 + 2\pi j \} - \max \{ a_1, a_2 + 2\pi i, a_3 + 2\pi j\} \big)^+ .
    \]
\end{lemma}

\begin{proof}
    Lift the intersection to~$[a_1,b_1]$. Up to endpoints, it is the
    disjoint union of
    \[
        [a_1,b_1]\cap[a_2+2\pi i,b_2+2\pi i]
        \cap[a_3+2\pi j,b_3+2\pi j],
        \qquad i,j\in\{-1,0,1\}.
    \]
    Each interval has length equal to the corresponding summand.
\end{proof}

\section{Omitted Proofs}
\label{sec:proofs}

\begin{lemma}[Formula for~$\rho$]

    \label{lem:rho}
    Let~$\sigma=\{\vect v,\vect v_1,\ldots,\vect v_k\}$,~$r=r_{\vect v}(t)\in(0,\pi)$, and suppose~$R$ has positive area.
    Set~$\vect n_i=N(\vect v-\vect v_i)$ and
    identify the unit circle in~$\vect c^\perp$ with~$\sph^1$.
    For~$\vect n_i$ not parallel to~$\vect c$, let~$I_i=B_{r_i}(\vect s_i)$, where
    \[
        r_i = \atantwo \Big( \sqrt{ \big( \sin^2 r - (\vect n_i \cdot \vect c)^2 \big)^+}, - \cos r (\vect n_i \cdot \vect c) \Big)
    \]
    and
    \[
        \vect s_i = N\big(\vect n_i-(\vect n_i\cdot\vect c)\vect c\big).
    \]
    For~$\vect n_i$ parallel to~$\vect c$, set~$I_i=\sph^1$ if~$\cos r\,(\vect n_i\cdot\vect c)\geq0$, and~$I_i=\emptyset$
    otherwise. Then
    \[
        \rho=\frac{1}{2\pi}\mu\left(\bigcap_{i=1}^k I_i\right),
    \]
    where~$\mu$ denotes arclength on~$\sph^1$. For~$k=0$, the
    intersection is~$\sph^1$ and~$\rho=1$. If any~$I_i$ is empty
    or has radius zero, then~$\rho=0$.
\end{lemma}

\begin{proof}[Proof of \lemref{rho}]

    Parametrize the cap boundary by~$\vect x=\cos r\,\vect c+\sin r\,\vect u$, where~$\vect u\in\sph^1\subset\vect c^\perp$. Writing~$d_i=\vect n_i\cdot\vect c$, the~$i$th hemisphere condition is
    \[
        \vect n_i\cdot\vect x
        =\cos r\,d_i+\sin r\,\vect n_i\cdot\vect u\geq0.
    \]
    If~$\vect n_i$ is parallel to~$\vect c$, this gives precisely the
    stated full-circle or empty-set case. Otherwise,~$\vect n_i\cdot\vect u=\sqrt{1-d_i^2}\,\vect s_i\cdot\vect u$,
    so the condition becomes
    \[
        \vect s_i\cdot\vect u\geq q_i,
        \qquad q_i=\frac{-\cos r\,d_i}{\sin r\sqrt{1-d_i^2}}.
    \]
    For~$|q_i|<1$, this defines a circular interval of radius~$\cos^{-1}(q_i)$. Because
    \[
        \sin^2 r(1-d_i^2)-\cos^2 r\,d_i^2
        =\sin^2 r-d_i^2,
    \]
    this radius equals the stated~$\atantwo$ expression.
    For~$q_i\leq-1$ the expression gives radius~$\pi$; for~$q_i\geq1$ it gives radius zero. Thus~$I_i$ represents the
    permitted directions up to sets of arclength zero.
    Intersecting the constraints gives~$\bigcap_i I_i$.
    The parametrization scales arclength by~$\sin r$, so
    \[
        \rho=\frac{\sin r\,\mu(\bigcap_i I_i)}{2\pi\sin r}
        =\frac{1}{2\pi}\mu\left(\bigcap_i I_i\right).
    \]

\end{proof}

\section{Additional Details On the Barley Seed Experiment}

\subsection{An Example Barley Seed}
\label{append:barley-seed}

\begin{figure}[htbp]
    \centering
    \includegraphics[width=\linewidth]{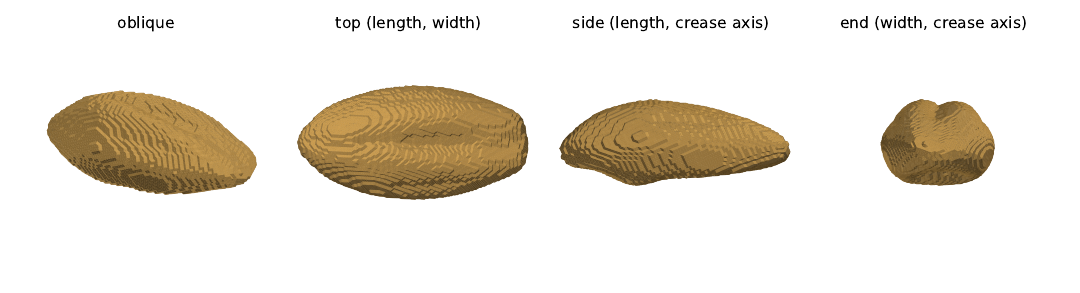}
    \caption{
        One seed of the barley dataset (type Algerian, 28,056 triangles), as the
        triangulated surface mesh obtained with marching cubes from its cleaned
        X-ray CT scan.
        The seed is shown in the hand-built frame of~\cite{amezquita2022measuring}:
        the first axis is the length of the seed, the second its width, and the
        third the axis on which the crease faces up.
        The steps on the surface arise from the voxels of the scan ($127\,\mu$m).
    }
    \label{fig:append:barley-seed}
\end{figure}

\figref{append:barley-seed} shows one seed from the barley dataset
of~\cite{amezquita2022measuring}, as the triangulated surface mesh on which we
compute the DIWECT and the WECT in \secref{exp:barley}.

\subsection{Full Barley Seed Classification Results}
\label{append:barley-table}

\tabref{exp:barley} lists the accuracies shown in \figref{exp:barley-pose},
together with those of the traditional descriptors of~\cite{amezquita2022measuring},
alone and combined with each topological feature (see \secref{exp:barley}).

\begin{table}[htbp]
    \centering
    \caption{
        Accuracy (mean~$\pm$ standard deviation over 20 splits) of classifying
        28 barley seed types.
        ``Flipped'' is the hand-built frame with random axis sign flips and
        width/thickness swaps.
        Chance is about~$0.04$.
        The DIWECT achieves comparable results to the WECT, while remaining
        robust to errors in the seed framing.
    }
    \label{tab:exp:barley}
    \begin{tabular}{lccc}
        \toprule
        features & dimension & seed split & spike split \\
        \midrule
        traditional descriptors & 11 &~$0.53 \pm 0.01$ &~$0.29 \pm 0.02$ \\
        \addlinespace
        DIWECT & 64 &~$0.39 \pm 0.01$ &~$0.24 \pm 0.02$ \\
        WECT, hand-built frame & 2,528 &~$\mathbf{0.45 \pm 0.01}$ &~$\mathbf{0.26 \pm 0.03}$ \\
        WECT, hand-built frame~$+ 2^\circ$ & 2,528 &~$0.39 \pm 0.02$ &~$0.24 \pm 0.02$ \\
        WECT, flipped & 2,528 &~$0.21 \pm 0.01$ &~$0.16 \pm 0.01$ \\
        WECT, automatic PCA & 2,528 &~$0.35 \pm 0.01$ &~$0.21 \pm 0.02$ \\
        WECT, scanner & 2,528 &~$0.32 \pm 0.02$ &~$0.12 \pm 0.01$ \\
        WECT, random rotation & 2,528 &~$0.10 \pm 0.01$ &~$0.09 \pm 0.01$ \\
        \addlinespace
        traditional~$+$ DIWECT & 75 &~$0.60 \pm 0.02$ &~$0.31 \pm 0.03$ \\
        traditional~$+$ WECT, hand-built frame & 2,539 &~$\mathbf{0.62 \pm 0.01}$ &~$\mathbf{0.33 \pm 0.04}$ \\
        traditional~$+$ WECT, flipped & 2,539 &~$0.44 \pm 0.01$ &~$0.25 \pm 0.03$ \\
        traditional~$+$ WECT, random rotation & 2,539 &~$0.39 \pm 0.01$ &~$0.24 \pm 0.02$ \\
        \bottomrule
    \end{tabular}
\end{table}

\subsection{Barley Seeds as Voxel Complexes}
\label{append:barley-voxels}

In \secref{exp:barley}, we compute the DIWECT of each barley seed's surface
mesh.
However, the ECT of~\cite{amezquita2022measuring} is computed on the cubical
complex of the voxels of each seed.
To check that our results do not depend on this choice of representation, we
also compute the DIWECT of the Freudenthal triangulation of each seed's cubical
complex, which equals the DIWECT of the cubical complex itself.
With the same classifier and splits as in \secref{exp:barley}, this DIWECT
achieves an accuracy of~$0.40$ on the seed split and~$0.22$ on the spike split,
compared with~$0.39$ and~$0.24$ for the DIWECT of the mesh.

\end{document}